\documentclass[prc,twocolumn,,superscriptaddress,showpacs,floatfix]{revtex4}

\usepackage{graphicx}
\usepackage{dcolumn}

\begin{document}               

\def\be{\begin{equation}}
\def\ee{\end{equation}}
\def\ba{\begin{eqnarray}}
\def\ea{\end{eqnarray}}
\def\bas{\begin{eqnarray*}}
\def\eas{\end{eqnarray*}}


\title{Solution of large scale nuclear structure problems by wave function
factorization}

\author{T.~Papenbrock}
\affiliation{Physics Division, 
Oak Ridge National Laboratory, Oak Ridge, TN 37831, USA}
\affiliation{Department of Physics and Astronomy, University of Tennessee,
Knoxville TN 37996-1201, USA}
\author{A. Juodagalvis}
\affiliation{Physics Division, 
Oak Ridge National Laboratory, Oak Ridge, TN 37831, USA}
\affiliation{Department of Physics and Astronomy, University of Tennessee,
Knoxville TN 37996-1201, USA}
\author{D.~J.~Dean}
\affiliation{Physics Division, 
Oak Ridge National Laboratory, Oak Ridge, TN 37831, USA}
\date{\today}

\begin{abstract}
Low-lying shell model states may be approximated accurately by a sum
over products of proton and neutron states. The optimal factors are
determined by a variational principle and result from the solution of
rather low-dimensional eigenvalue problems. Application of this method
to $sd$-shell nuclei, $pf$-shell nuclei, and to no-core shell model
problems shows that very accurate approximations to the exact
solutions may be obtained. Their energies, quantum numbers and
overlaps with exact eigenstates converge exponentially fast as the
number of retained factors is increased. 
\end{abstract}


\maketitle 

\section{Introduction}

Realistic nuclear structure models are difficult to solve due to the
complexity of the nucleon-nucleon interaction and the sheer size of
the model spaces. Exact diagonalizations are now possible for
$pf$-shell nuclei \cite{Antoine,Nathan,Honma02} and for sufficiently
light systems \cite{Navratil00,Navratil02}, and Quantum Monte Carlo
calculations \cite{Pieper01,Pieper02} have solved light nuclei up to
about mass $A = 12$.  For cases where an exact solution is not
feasible, various approximations are employed. We mention stochastic
methods like shell-model Monte Carlo \cite{Lang93,Koonin97} and Monte
Carlo shell-model \cite{MCSM}, and recent applications of coupled
cluster expansions \cite{Heisenberg,DeanCC}. 

In recent years, several truncation method for shell model
diagonalizations have been developed. These methods aim at a reduction
of the enormous dimensionality of shell model Hilbert spaces while
maintaining a high accuracy in the computed observables. Based on
arguments of statistical spectroscopy, Horoi and
coworkers~\cite{Horoi94,Horoi99,Horoi02,Horoi03} developed the
exponential convergence method. Mizusaki and
Imada~\cite{Mizu02,Mizu03} devised an extrapolation method and applied
it to a configuration truncation.  Both techniques use single-particle
basis states and provide a method to extrapolate the results of
truncated calculations to the full Hilbert space. Other approaches use
correlated basis states to obtain a rapid convergence. Andreozzi {\it
et al.}  \cite{AP,ALP}, for instance, construct a basis from products
of correlated proton states and correlated neutron states.  Gueorguiev
{\it et al.}~\cite{Vesselin01,Vesselin02} use a mixed-mode shell model
of single-particle and collective configurations, while Vargas {\it et
al.}~\cite{VHD} use a truncation based on coupled $SU(3)$ irreps to
describe the interplay and competition of collective and
single-particle degrees of freedom.

Though the selection of the relevant states is physically well
motivated for all these truncation schemes, it does not directly
follow from a variational principle. This is different for the density
matrix renormalization group (DMRG)~\cite{White92} and the very
recently proposed factorization method~\cite{PD03}. The DMRG uses a
sophisticated renormalization and truncation scheme that includes the
most important states and correlations. Dukelsky {\it et
al.}~\cite{Duk01,Duk02} and Dimitrova {\it et al.}~\cite{Dimitrova02}
applied this method to nuclear structure problems.  The ground-state
factorization is based on a related truncation.  At a given
truncation, the optimal states are determined from a variational
principle. These last two methods also allow for an extrapolation to
full Hilbert spaces as the results tend to converge exponentially. In
this article we present a detailed description of the factorization
method and discuss several applications. A summary of some of the main
results has been presented in an earlier paper \cite{PD03}.

This article is organized as follows. In Section~\ref{theory} we give
a derivation of the main theoretical results and present details of
the numerical implementation. Section~\ref{numerics} presents
numerical results for $sd$-shell nuclei, $pf$-shell nuclei and for
no-core shell-model problems. The convergence of the factorization
method and a comparison with other truncation methods is presented in
Sect.~\ref{conv}, and we conclude with Section~\ref{concl}.

\section{Theoretical background}
\label{theory}

\subsection{Motivation}

Shell model basis states are products of proton Slater determinants
$\{|\pi_\alpha\rangle,\alpha=1,\ldots,d_P\}$ and neutron Slater
determinants $\{|\nu_\alpha\rangle,\alpha=1,\ldots,d_N\}$. Here, $d_P$
and $d_N$ denote the dimension of the proton space and the neutron
space, respectively. The shell-model ground-state may be expanded as
\be
\label{state}
|\Psi\rangle=\sum_{\alpha}^{d_P}\sum_{\beta}^{d_N} 
\Psi_{\alpha \beta}
|\pi_\alpha\rangle|\nu_\beta\rangle.
\ee 
This expansion is not unique since the amplitudes $\Psi_{\alpha
\beta}$ depend on the choice of basis states within the two subsets.
There is, however, a preferred basis in which the amplitudes
$\Psi_{\alpha \beta}$ are ``diagonal''. This basis is formally
obtained from a singular value decomposition of the rectangular
amplitude matrix $\Psi$ of dimension $d_P\times d_N$ as $\Psi= U S
V^T$. Here $U$ ($V$) is a $d_P\times d_P$ ($d_N\times d_N$)
dimensional matrix with orthonormalized columns. $S$ is a rectangular 
matrix of dimension $d_P\times d_N$ with elements $S_{i,j}=0$ for $i\ne j$ and 
the ``diagonal'' elements $S_{i,i}=s_i$ are
the non-negative singular values. Performing the singular value
decomposition yields
\be
\label{svd}
|\Psi\rangle=\sum_{j=1}^{\min{\left(d_P,d_N\right)}}
s_j |\tilde{p}_j\rangle|\tilde{n}_j\rangle.
\ee
Here $s_j$ denote the singular values while the proton-states
$|\tilde{p}_j\rangle$ and the neutron-states $|\tilde{n}_j\rangle$ are
orthonormal sets of left and right singular vectors, respectively.  In
general, these states are superpositions of many Slater determinants
and exhibit strong correlations.  The non-negative singular values
$s_1\ge s_2\ge s_3 \ldots$ fulfill $\sum_j s_j^2=1$ due to wave
function normalization.

\begin{figure}[t]
\vskip 0.3cm
\includegraphics[width=0.4\textwidth]{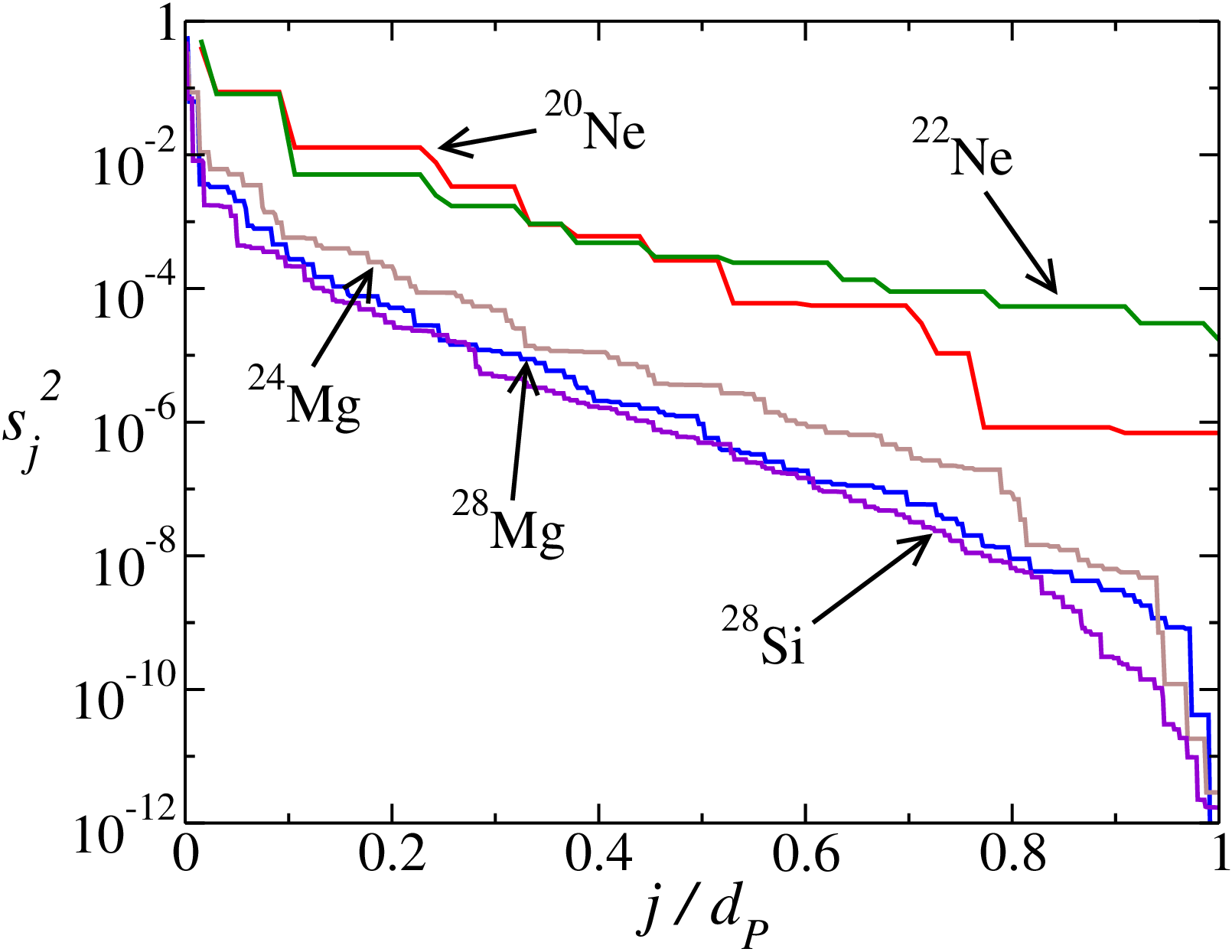}
\caption{\label{fig1}Singular values $s_j^2$ for ground states of 
${}^{20}$Ne ($d_{P}=66$),
${}^{22}$Ne ($d_{P}=66$),
${}^{24}$Mg ($d_{P}=495$),
${}^{28}$Mg ($d_{P}=495$), and
${}^{28}$Si ($d_{P}=924$).}
\end{figure}

It is interesting to compute the singular value decomposition for
ground states of realistic nuclear many-body Hamiltonians. To this
purpose we perform a numerical singular value decomposition of the
amplitude matrix $\Psi_{\alpha \beta}$ in Eq.~(\ref{state}) using the
{\sc Lapack} routines \cite{lapack}.  Figure~\ref{fig1} shows the
squares of the singular values for the ground states of $sd$-shell
nuclei ${}^{20}$Ne, ${}^{22}$Ne, ${}^{24}$Mg and $^{28}$Si (from the
USD interaction \cite{BW}) plotted versus their normalized index
$j/d_P$.  (We have $d_P=190, 190, 495$, and 924 for the nuclei
${}^{20}$Ne, ${}^{22}$Ne, ${}^{24}$Mg and $^{28}$Si,
respectively). The singular values decrease with increasing index and
rapidly become exponentially small. (Degeneracies are due to
spin/isospin symmetry).  This suggests that a truncation of the sum in
Eq.~(\ref{svd}) should yield an accurate approximation to the
ground-state. In fact, the density matrix renormalization group (DMRG)
\cite{White92}, exploits this rapid fall-off of singular values in a
wave-function factorization. For obvious reasons, the expansion
(\ref{svd}) is a factorization, and the correlated proton and neutron
states are the factors.  In what follows, we will devise a method that
directly obtains the most important factors without knowledge of the
exact ground-state.

\subsection{Derivation of main results}

To determine the optimal factor for a given truncation we make the ansatz
\be
\label{ansatz}
|\psi\rangle=\sum_{j=1}^\Omega |p_j\rangle|n_j\rangle
\ee
for the ground state.
Here, the unknown factors are the proton-states $|p_j\rangle$ and
the neutron-states $|n_j\rangle$. These states may be correlated and need 
not to be normalized. The truncation is controlled by
the parameter $\Omega$ which counts the number of desired
factors. Figure~\ref{fig1} suggests that $\Omega\ll\min{\left(d_P,d_N\right)}$ 
yields accurate approximations to shell-model
ground states. This is also the result of our numerical computations
below.

Let $\hat H$ be the nuclear many-body Hamiltonian.
To determine the unknown proton-states $|p_j\rangle$ and neutron-states
$|n_j\rangle$ in Eq.~(\ref{ansatz}) we consider the 
energy $E=\langle\psi|\hat{H}|\psi\rangle/\langle\psi|\psi\rangle$. Its
variation $\delta E=0$ yields ($j=1,\ldots,\Omega$)
\ba
\label{solution}
\sum_{i=1}^\Omega\left(\langle n_j|\hat{H}|n_i\rangle 
  - E\langle n_j|n_i\rangle\right)|p_i\rangle &=& 0,\nonumber\\
\sum_{i=1}^\Omega\left(\langle p_j|\hat{H}|p_i\rangle 
  - E\langle p_j|p_i\rangle\right)|n_i\rangle &=& 0.
\ea
The solution of these nonlinear equations determines the optimal factors
and the ground-state energy. Note
that for fixed neutron (proton) states the first (second) set
of these equations constitutes a generalized eigenvalue problem for
the proton (neutron) states.  To fully understand the structure of the matrices
involved we rewrite the
first set of the Eq.~(\ref{solution}) as
\begin{widetext}
\ba
\label{matrixform}
\left(\begin{array}{cccc}
   \langle n_1|\hat{H}|n_1\rangle & \langle n_1|\hat{H}|n_2\rangle &\cdots &
\langle n_1|\hat{H}|n_\Omega\rangle \\
\langle n_2|\hat{H}|n_1\rangle &\langle n_2|\hat{H}|n_2\rangle &\cdots &
\langle n_2|\hat{H}|n_\Omega\rangle \\
\vdots &   & \ddots & \vdots \\
\langle n_\Omega|\hat{H}|n_1\rangle &\cdots & \cdots &
                  \langle n_\Omega|\hat{H}|n_\Omega\rangle \end{array}\right)
\left( \begin{array}{c}
   |p_1\rangle \\
   |p_2\rangle \\
   \vdots   \\
   |p_\Omega\rangle \end{array} \right)
= E
\left(\begin{array}{cccc}
   \langle n_1|n_1\rangle {\hat I}_{P}&
\langle n_1|n_2\rangle {\hat I}_{P}&\cdots &
\langle n_1|n_\Omega\rangle {\hat I}_{P}\\
\langle n_2|n_1\rangle {\hat I}_{P}&
\langle n_2|n_2\rangle {\hat I}_{P}&\cdots &
\langle n_2|n_\Omega\rangle {\hat I}_{P}\\
\vdots &  & \ddots & \vdots \\
\langle n_\Omega|n_1\rangle {\hat I}_{P}&\cdots & \cdots &
\langle n_\Omega|n_\Omega\rangle{\hat I}_{P}
\end{array}\right)
\left(\begin{array}{c}
   |p_1\rangle \\
   |p_2\rangle \\
   \vdots   \\
   |p_\Omega\rangle \end{array}\right).
\ea
\end{widetext}
Here, $|p_j\rangle=(p_{j,1},p_{j,2},\ldots,p_{j,d_P})^T$ is a
$d_P$-dimensional column vector ($^T$ denotes the transpose) while
${\hat I}_{P}$ denotes the identity matrix in proton space. Thus,
\ba 
\langle n_i|n_j\rangle {\hat I}_{P}=\left(\begin{array}{cccc} 
\langle n_i|n_j\rangle &0 &\cdots &0\\ 
0 & \langle n_i|n_j\rangle & & 0\\ 
\vdots & & \ddots & \vdots \\ 
0& \cdots & \cdots & \langle n_i|n_j\rangle
\end{array}\right)
\ea
is a diagonal constant matrix of dimension $d_P\times d_P$.
The proton-space operators $\langle n_i|\hat{H}|n_j\rangle$ 
stem from the nuclear structure Hamiltonian
\be
\label{ham}
\hat{H}=\hat{H}_{N} + \hat{H}_{P} +\hat{V}_{PN}, 
\ee
with
\ba
\hat{H}_{N}&=&\sum_n \varepsilon_n \hat{a}_n^\dagger\hat{a}_n
+ {1\over 4}\sum_{n,n',m,m'}v_{n n' m' m}\hat{a}_n^\dagger
\hat{a}^\dagger_{n'}\hat{a}_m\hat{a}_{m'},\nonumber\\
\hat{H}_{P}&=&\sum_p \varepsilon_p \hat{a}_p^\dagger\hat{a}_p
+ {1\over 4}\sum_{p,p',q,q'}v_{p p' q' q}\hat{a}_p^\dagger
\hat{a}^\dagger_{p'}\hat{a}_q\hat{a}_{q'},\\
\hat{V}_{PN}&=& \sum_{p,n,n',p'} v_{p n p' n'}
\hat{a}_p^\dagger\hat{a}^\dagger_{n}\hat{a}_{n'}\hat{a}_{p'}.\nonumber
\ea
Here, we use indices $p,q$ and $m,n$ to refer to proton and neutron
orbitals, respectively. The antisymmetric two-body matrix elements are
denoted as $v_{i j k l}$.

Thus, the proton-space Hamilton operator 
$\langle n_i|\hat{H}|n_j\rangle$ is 
\ba
\label{Hpp}
\langle n_i|\hat{H}|n_j\rangle &=& 
\sum_{p,p'} \left(\sum_{n,n'}v_{p n p' n'}
\langle n_i|\hat{a}^\dagger_{n}\hat{a}_{n'}|n_j\rangle\right)
\hat{a}_p^\dagger\hat{a}_{p'}\nonumber\\
&+& \langle n_i|\hat{H}_{N}|n_j\rangle+
\langle n_i|n_j\rangle\hat{H}_{P}. 
\ea
Note that the neutron-proton interaction $\hat{V}_{PN}$ results into
a one-body proton operator while the neutron Hamiltonian
$\hat{H}_{N}$ yields a constant. This concludes the detailed 
explanation of the first set of equations in Eq.~(\ref{solution}). 
The second set has an identical structure, only the role of neutrons and
protons is reversed.

\subsection{Treatment of symmetries}

Most modern shell-model codes use a basis of Slater
determinants that preserve axial symmetry. It is straight forward to
include this symmetry into the method proposed in this work. For such
an ``$m$-scheme'' ground-state factorization we modify the ansatz
(\ref{ansatz}) as
\be
\label{mscheme}
|\psi\rangle=\sum_{m=-M}^M\sum_{k=1}^{\Omega_m} |p_k(m)\rangle
|n_k(-m)\rangle.
\ee
Here $|p_k(m)\rangle, k=1,\ldots,\Omega_m$ ($|n_k(-m)\rangle,
k=1,\ldots,\Omega_{m}$) denote $d_{P,m}$ dimensional proton
states ( $d_{N,-m}$ dimensional neutron states) with angular
momentum projection $J_z=m$ ($J_z=-m$), and the sum over $m$
runs over all possible values of $J_z$. The ansatz (\ref{mscheme})
leads to a generalized eigenvalue problem similar to
Eq.~(\ref{solution}), with the only difference that permissible
products of proton states and neutron states have zero angular
momentum projection: 
\ba
\label{msolp}
\lefteqn{ \sum_{m=-M}^M \sum_{k=1}^{\Omega_m} \bigg( 
\langle n_{k'}(m')|\hat{H}|n_k(m)\rangle } \nonumber\\ 
&& - E\langle n_{k'}(m')|n_k(m)\rangle \bigg) |p_k(-m)\rangle = 0, 
\ea
\ba
\label{msoln}
\lefteqn{ \sum_{m=-M}^M \sum_{k=1}^{\Omega_m} \bigg( 
\langle p_{k'}(m')|\hat{H}|p_k(m)\rangle }\nonumber\\ 
&& - E\langle p_{k'}(m')|p_k(m)\rangle \bigg) |n_k(-m)\rangle = 0. 
\ea
These equations have to be fulfilled for all possible values of $m'$ and $k'$.
It is again useful to display the eigenvalue problem~(\ref{msolp}) 
for the proton states in more detail
\begin{widetext}
\ba
\label{mmatrix}
\left(\begin{array}{ccc}
   \langle n(-M)|\hat{H}|n(-M)\rangle &\cdots &
\langle n(-M)|\hat{H}|n(M)\rangle \\
\vdots & \ddots & \vdots \\
\langle n(M)|\hat{H}|n(-M)\rangle & \cdots &
                  \langle n(M)|\hat{H}|n(M)\rangle \end{array}\right)
\left( \begin{array}{c}
   |p(M)\rangle \\
   \vdots   \\
   |p(-M)\rangle \end{array} \right)
= E
\left(\begin{array}{c}
   \langle n(-M)|n(-M)\rangle\quad|p(M)\rangle \\
   \vdots   \\
   \langle n(M)|n(M)\rangle\quad|p(-M)\rangle \end{array}\right).
\ea
\end{widetext}
Here, we used the shorthands  
\ba 
|p(m)\rangle&=&(|p_1(m)\rangle,|p_2(m)\rangle,\cdots,
|p_{\Omega_m}(m)\rangle)^T,
\ea  
the rectangular block matrices
\bas
\lefteqn{ \langle n(k)|\hat{H}|n(l)\rangle =} \\ 
& & \left(\begin{array}{ccc}
   \langle n_1(k)|\hat{H}|n_1(l)\rangle &\cdots &
\langle n_1(k)|\hat{H}|n_{\Omega_l}(l)\rangle \\
\vdots & \ddots & \vdots \\
\langle n_{\Omega_k}(k)|\hat{H}|n_1(l)\rangle & \cdots &
   \langle n_{\Omega_k}(k)|\hat{H}|n_{\Omega_l}(l)\rangle \end{array}\right),
\eas
and the overlap matrices
\bas
\lefteqn{ \langle n(k)|n(k)\rangle = }\\
& & \left(\begin{array}{ccc}
   \langle n_1(k)|n_1(k)\rangle \hat{I}_{P,-k}&\cdots &
\langle n_1(k)|n_{\Omega_k}(k)\rangle \hat{I}_{P,-k}\\
\vdots & \ddots & \vdots \\
\langle n_{\Omega_k}(k)|n_1(k)\rangle \hat{I}_{P,-k}& \cdots &
   \langle n_{\Omega_k}(k)|n_{\Omega_k}(k)\rangle \hat{I}_{P,-k}
\end{array}\right).
\eas
Note that the right hand side of Eq.~(\ref{mmatrix}) is a vector.
Note also that $\langle n_i(k)|\hat{H}|n_j(l)\rangle$ is a rectangular
$d_{P,k}\times d_{P,l}$ matrix similar to Eq.~(\ref{Hpp}), while
$\hat{I}_{P,k}$ denotes the identity operator for the proton sub-space
with angular momentum $J_z=k$. The eigenvalue
problem~(\ref{mmatrix}) differs from the eigenvalue
problem~(\ref{matrixform}) due to the block diagonal overlap matrix on
its right hand sight.  The eigenvalue problem for the neutrons is
identical to Eq.~(\ref{mmatrix}) when the role of neutrons and protons
is reversed.

The number of factors used in the
$m$-scheme factorization is given by the parameters $\Omega_m$. These
parameters are input to the factorization. In the following, we use
\be
\label{frac}
\Omega_m=\Omega_m(\alpha)=\max{\left(1, \alpha d_{P,m}\right)},
\ee
and recall that $d_{P,m}$ is the dimension of the proton-subspace with
angular momentum projection $J_z=m$. For $\alpha=0$ the most severe
truncation is obtained and leads us to solve eigenvalue problems of
the dimension $d_P$ and $d_N$, respectively. Setting $\alpha=1$ leads
to an eigenvalue problem with the same dimension as an exact
diagonalization in $m$-scheme. Below we will see that the
choice~(\ref{frac}) of parameters yields rapidly converging
results. However, there may be other parameterizations that are
superior. Comparison of Eq.~(\ref{mmatrix}) and Eq.~(\ref{matrixform})
shows that the $m$-scheme factorization yields a lower-dimensional
eigenvalue problem than the factorization (\ref{ansatz}) if the same
number of factors is used.

Note also that a $jj$-coupled scheme can be used. Let
$|p_k(J,m)\rangle$ ($|n_k(J,m)\rangle$) be a proton (neutron) state
with angular momentum quantum number $J$ and angular momentum
projection $J_z=m$. A ground-state of an even-even nucleus has $J=0$ and 
can be factored as
\be
\label{jj}
|\psi\rangle=\sum_J\sum_{m=-J}^J\sum_{k=1}^{\Omega_{J,m}} |p_k(J,m)\rangle
|n_k(J,-m)\rangle,
\ee
where $\Omega_{J,m}$ is the number of retained states with angular momentum 
quantum number $J$ and angular momentum projection $J_z=m$. We choose
\be
\label{fracj}
\Omega_{J,m}(\alpha)=\max{\left(1, \alpha d_{P,J,m}\right)},
\ee
where $d_{P,J,m}$ is the dimension of the proton-subspace
with angular momentum $J$ and projection $m$. This number is independent
of $m$ for fixed $J$. Generalizations of the ansatz (\ref{jj}) to  nonzero 
$J$ are straight forward. 

Other symmetries like parity can also be used to further reduce the
dimensionality of the eigenvalue problem. Parity even states, for
instance, are products of parity even proton states with parity even
neutron states or products off parity odd proton states with parity odd
neutron states. The ability to implement symmetries is particularly
important as it widens the flexibility of the ground-state
factorization. Consider for instance shell-model problems with proton
and neutron spaces that differ considerably in size such that $d_P \ll
d_N$. In such cases one might switch to a more symmetric factorization
and factorize the ground-state as follows: one of the factor spaces
would consist of neutron states that are based on a subset of neutron
single-particle orbitals, while the other factor states would be based
on the remaining neutron orbitals and all proton orbitals. In such a
scenario, the correct implementation of the Pauli principle requires
care.

\subsection{Numerical solution and computational details}

We discuss the solution of the eigenvalue problems. For notational
convenience we focus on the solution of the equations
(\ref{solution}). The corresponding equations (\ref{msolp}) and
(\ref{msoln}) of the $m$-scheme factorization or the can be treated
similarly. We solve the coupled set of nonlinear equations
(\ref{solution}) in an iterative procedure. We choose a random set of
linearly independent initial neutron-states $\{|n_j\rangle,
j=1,\ldots,\Omega\}$ and construct the Hamiltonian and overlap matrix
presented in Eq.~(\ref{matrixform}). We then solve this generalized eigenvalue
problem of dimension $\Omega d_P$ for those
proton-states $\{|p_j\rangle,j=1,\ldots,\Omega\}$ that yield the
lowest energy $E$. In practice, we use the sparse matrix solver {\sc
Arpack} \cite{arpack} for this task. The solution of the generalized
eigenvalue problem requires us to provide the LU factorization of the
overlap matrix (i.e. the right hand side of
Eq.~(\ref{matrixform})). This factorization simplifies considerably
since the overlap matrix is a direct product $\hat{I}_P\otimes M_N$
and only requires the LU-factorization of the $\Omega\times\Omega$
dimensional overlap matrix $M_N$ with elements $\langle n_i |
n_j\rangle$. We then input the resulting proton-states to the second
set of Eq.~(\ref{solution}). The solution of this $\Omega d_N$
dimensional problem yields improved neutron-states
$\{|n_j\rangle,j=1,\ldots,\Omega\}$ and an improved (i.e. lowered)
ground-state energy $E$. We iterate this procedure until the
ground-state energy $E$ is converged.

For small values of $\Omega$ we typically need about 20 iterations to
obtain a converged energy, and the number of iterations decreases with
increasing number of kept states $\Omega$. For maximal value
$\Omega=d_N$ , the first solution of the proton-problem (\ref{matrixform})
already yields the exact ground-state. This can be seen as follows:
Using Slater determinants $\{|n_j\rangle=|\nu_j\rangle, j=1,\ldots,\Omega\}$
as input to the proton-problem (\ref{matrixform}) yields a matrix-problem
that is identical in structure to the full shell-model problem. 

Let us compare the effort of the ground-state factorization with an
exact diagonalization. Both methods require all Hamiltonian matrix
elements $\langle \pi_\alpha|\langle\nu_\beta|
\hat{H}|\nu_\gamma\rangle|\pi_\delta\rangle$. The
advantage of the factorization method is that the dimensionality of
the eigenvalue problem is $\Omega\times\max{(d_{P},d_{N})}$ with
$\Omega \ll d_{P}, d_{N}$ while the dimension of the exact diagonalization
scales like $d_{P}\times d_{N}$. Note that existing shell-model codes may
easily be modified to include the ground-state factorization in
order to obtain accurate approximations or to compute useful starting
vectors for a Lanczos iteration.

It is possible to reduce the Eqs.~({\ref{solution}) to a standard
eigenvalue problem. For this purpose we choose a random orthonormal set
of initial neutron-states $\{|n_j\rangle,j=1,\ldots,\Omega\}$ as input
to the first eigenvalue problem. This reduces the ``overlap'' matrix
$\langle n_j|n_i\rangle$ to a unit matrix, and we solve a standard
eigenvalue problem to obtain the proton-states
$\{|p_j\rangle,j=1,\ldots,\Omega\}$. The resulting proton-states will
not be orthogonal since they are components of only one solution
vector of dimension $\Omega d_P$. Their coefficient matrix $C$ with
elements $C_{\alpha j}\equiv \langle \pi_\alpha | p_j\rangle$ may,
however, be factorized in a singular value decomposition as $C=U D
V^T$. Here $D$ denotes a diagonal $\Omega\times\Omega$ matrix while
$U$ is a (column) orthogonal $d_{P} \times\Omega$ matrix, and $V$
is a orthogonal $\Omega\times\Omega$ matrix. The transformed states
($j=1,\ldots,\Omega$) 
\bas 
|p_j'\rangle&=&\sum_{\alpha=1}^{d_{P}}
U_{\alpha,j}|\pi_\alpha\rangle, \nonumber\\
|n_j'\rangle&=&\sum_{i=1}^\Omega V_{ij} |n_i\rangle 
\eas 
are orthonormal in the proton-space and in the neutron space, respectively,
and they fulfill 
\be 
|\psi\rangle=\sum_{j=1}^\Omega |p_j\rangle|n_j\rangle
= \sum_{j=1}^\Omega D_{jj}|p_j'\rangle|n_j'\rangle.  
\ee 

The orthogonal proton-states $\{|p_j'\rangle,j=1,\ldots,\Omega\}$ are
then input to the second set in Eq.~(\ref{solution}), which poses a
standard eigenvalue problem for the neutron-states. Note that the
transformed neutron states $D_{jj}|n'_j\rangle$ are useful starting
vectors for the Lanczos iteration of the sparse matrix solver. The
resulting neutron-states should again be orthonormalized by a singular
value decomposition.  These singular value decompositions are very
inexpensive compared to the diagonalization. 
Similarly, we may cast the generalized $m$-scheme eigenvalue problem
(\ref{mmatrix}) into a standard eigenvalue problem by enforcing
orthogonality $\langle n_k(m)|n_l(m)\rangle\propto\delta_{k,l}$
between neutron states with identical angular momentum projection
through the singular value decomposition.

In our implementation of the $m$-scheme factorization, we compute the
matrices of the proton space operators $H_P$ and
$\hat{a}_p^\dagger \hat{a}_{p'}$ as well as the matrices of the neutron
space operators $H_N$ and $\hat{a}_n^\dagger \hat{a}_{n'}$. These
sparse matrices can be stored in fast memory. The sparse matrix on the
left hand side of Eq.~(\ref{matrixform}) is constructed from these
matrices. This sparse matrix can be kept in fast memory for sufficiently small
problems; for larger problems, its nonzero matrix elements along with
the row-column information can be stored in large chunks of data on
disk without a severe increase of computing times. 

Our implementation of the $jj$-coupled factorization is somewhat more
tedious. Starting from the $m$-scheme basis states $|p_k(m)\rangle$ and
$|n_k(m)\rangle$, we create basis states $|p_k(J,m)\rangle$ and
$|n_k(J,m)\rangle$ with good angular momentum by numerical projection.
The projection operator is \cite{Lowdin}
\be
\hat{P}_J=\prod_{j\ne J} \frac{\hat{J}^2-j(j+1)}{J(J+1)-j(j+1)},
\ee
where the product over $j$ runs over all possible angular momenta, and
$\hat{J}$ denotes the angular momentum operator.  The matrices of the
proton Hamiltonian $H_P$ and the neutron Hamiltonian $H_N$ are then
transformed to this basis and stored. The matrices of the operators
$\hat{a}_p^\dagger \hat{a}_{p'}$ and $\hat{a}_n^\dagger \hat{a}_{n'}$
are not sparse in the basis with good $J$, and are therefore kept in the
$m$-scheme basis. We perform the appropriate basis
transforms in the construction of the Hamiltonian matrices.

\section{Numerical results}
\label{numerics}

A successful application of the ground state factorization would yield
accurate approximations from calculations involving rather small
dimensional eigenvalue problems. Clearly, the outcome depends on the
model space and the interaction under consideration. In this section
we apply the $m$-scheme factorization to realistic structure
calculations involving $sd$-shell nuclei, $pf$-shell nuclei and
no-core shell-model computations for $^4$He. The results are compared
to exact diagonalizations. We expect the method to converge most
rapidly in the case of weak proton-neutron correlations.  Thus, the
study of $N=Z$ nuclei provides a challenging testing
ground since $T=0$ proton-neutron correlations may be 
strong in these systems. Throughout this 
section $d$ denotes the dimension of the
$m$-scheme eigenvalue problems (\ref{msolp}) and (\ref{msoln}) we
actually solve, while $d_{\rm max}$ denotes the $m$-scheme dimension
required for an exact solution of the problem.

\subsection{$sd$-shell nuclei}

We apply the $m$-scheme factorization to the $sd$-shell nuclei
$^{24}$Mg, $^{26}$Al, $^{28}$Mg, and $^{28}$Al and use the USD
interaction. While the factorization is particularly suited to compute
accurate approximations of the ground states, we may also use it for
the computation of low-lying excited states. In some cases, excited
states can be obtained as a by-product of the ground-state
computation.  While solving the eigenvalue problems (\ref{msolp})
and (\ref{msoln})
for the ground-state, we may also compute excited state solutions. Let
us assume that we solve the equations (\ref{msolp})
for the proton states. The excited proton state solutions will be
obtained in the presence of neutron states that are optimized for the
ground state.  Therefore, we expect that the excited states are less
accurately reproduced than the ground-state. Figure~\ref{fig2} 
shows the resulting low-energy spectra for $^{24}$Mg,
$^{28}$Mg, $^{26}$Al, and $^{28}$Al.

\begin{figure}[t]
\vskip 0.3cm
\includegraphics[width=0.23\textwidth]{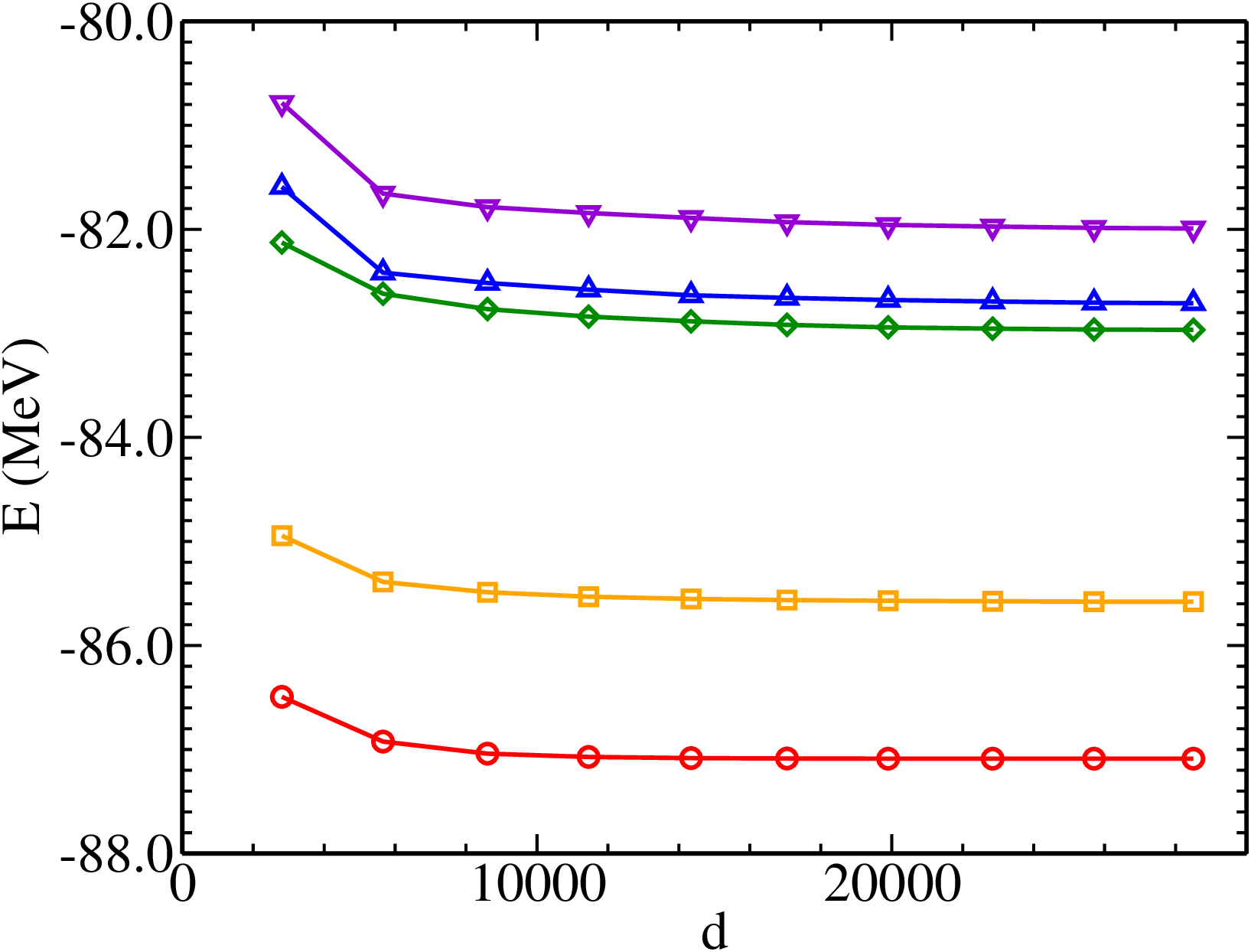}
\includegraphics[width=0.23\textwidth]{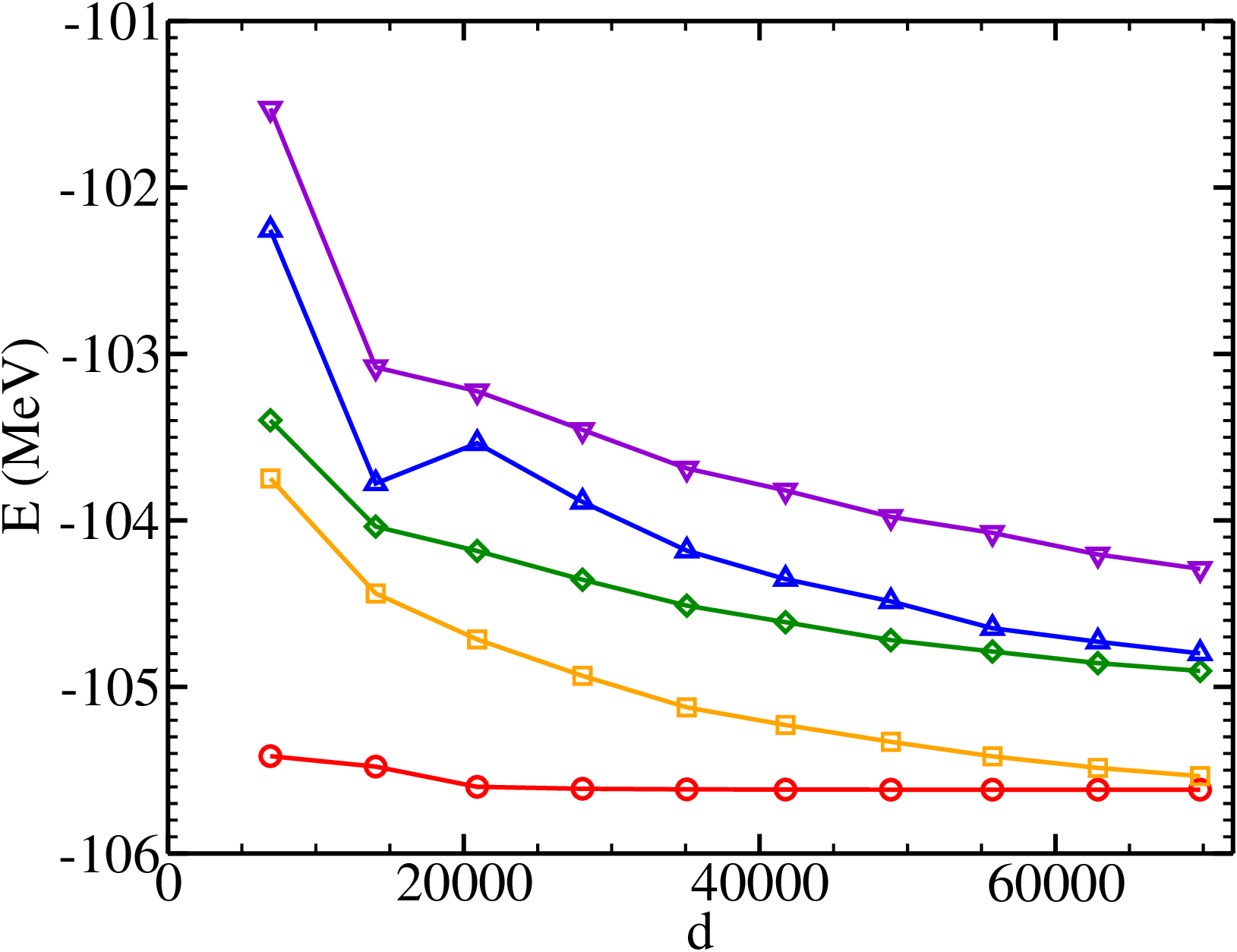}
\includegraphics[width=0.23\textwidth]{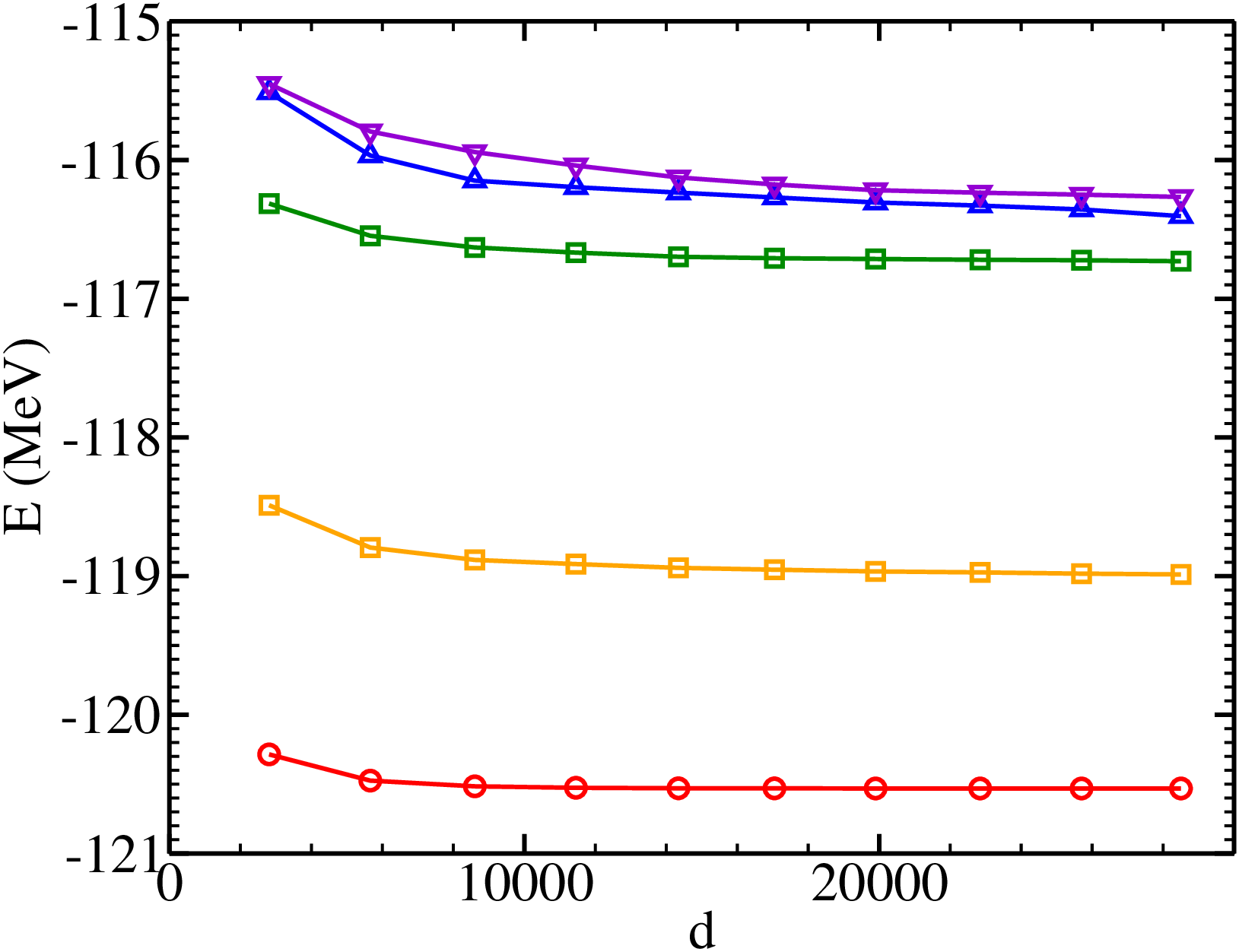}
\includegraphics[width=0.23\textwidth]{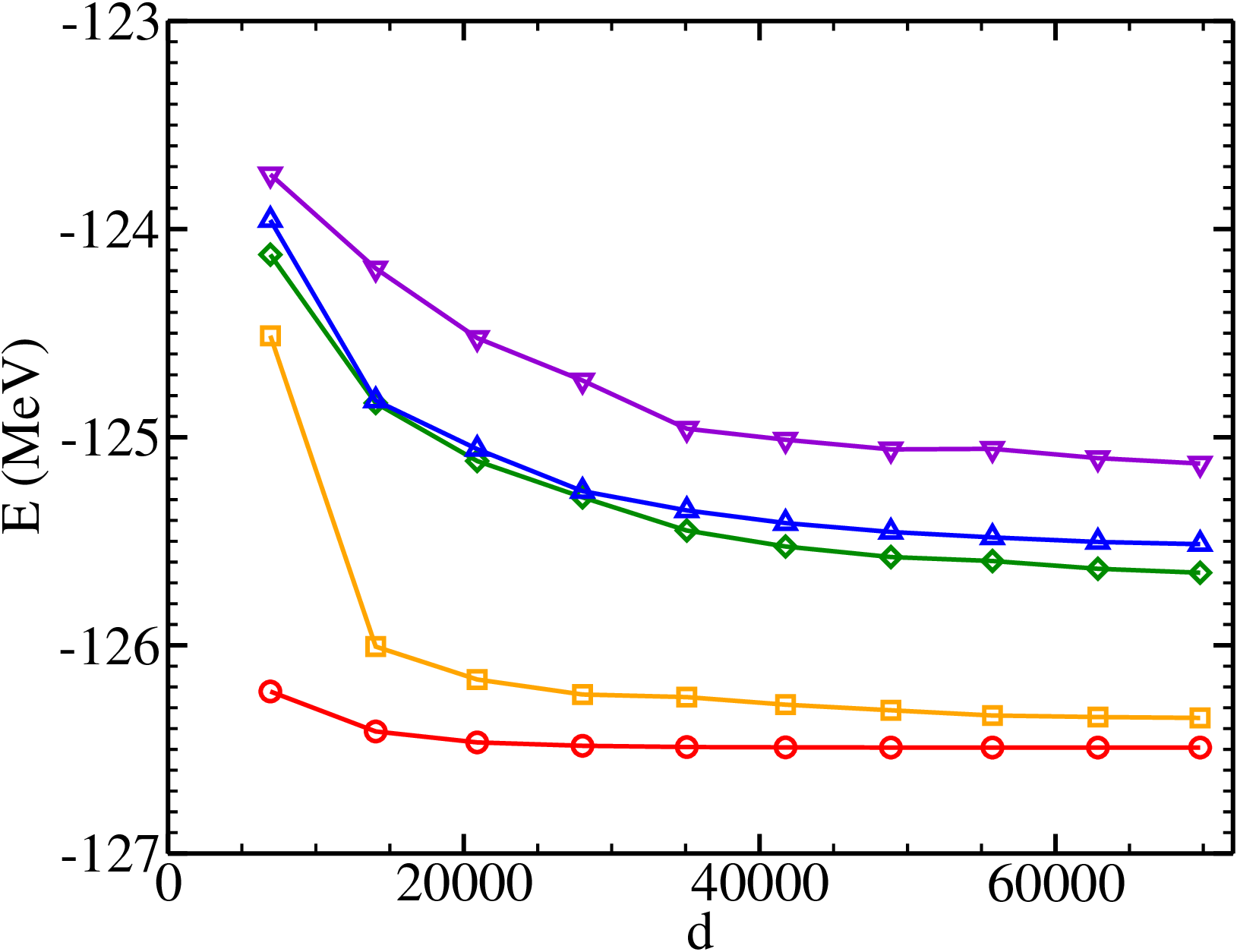}
\caption{\label{fig2}Low-energy spectrum of $sd$-shell 
nuclei versus the dimension $d$ of the eigenvalue problem. Results
are obtained from targeting the ground state. Left panel:
Mg isotopes ($d_{\rm max}=28503$) 
${}^{28}$Mg (top), ${}^{24}$Mg (bottom). Right panel: 
Al isotopes ($d_{\rm max}=69784$) ${}^{26}$Al (top)
${}^{28}$Al (bottom).}
\end{figure}

The ground states converge most rapidly as more factors are retained
in the factorization and the dimension $d$ of the eigenvalue problem
increases. Typically, excellent results are obtained from computations
involving relative dimensions $d/d_{\rm max}\approx 1/4\ldots 1/3$.
For the Mg isotopes, excited states converge somewhat slower than the
ground states, and level spacings are reproduced to a very good
accuracy already at severe truncations. This shows that the factors of
the low-lying excitations have a large overlap with the corresponding
ground state factors, and we assume that this finding is related to
the band structure of the low-lying excitations.  The situation is
different for the Al isotopes, as evident from the right panel of
Fig.~\ref{fig2}.  This slower convergence of excited states is not
unexpected due to the absence of band structure in these nuclei.

There are at least two approaches to improve the convergence of the
excited states. In the first approach, we may target excited states by
solving the eigenvalue problems~(\ref{msolp}) and (\ref{msoln})
directly for an excited state. This method works well for the first
and second excited states of $^{24}$Mg, as shown in
Fig.~\ref{fig3}. However, this approach is unstable for higher excited
states of $^{24}$Mg and for the first excited state of $^{26}$Al, as
the solutions of the eigenvalue problem are oscillating but fail to
converge with increasing number of iterations. In the case of
$^{26}$Al we proceed as follows. The ground state has angular momentum
$J=5$ while the first excited state has angular momentum $J=0$. We may
thus use the $jj$-coupled ansatz~(\ref{jj}) to compute the
lowest-lying state with angular momentum $J=0$. This yields the first
excited state. Figure~\ref{fig4} shows the results plotted versus the
corresponding $m$-scheme dimension. The convergence is much improved
and comparable to that of the ground-state.

\begin{figure}[t]
\vskip 0.3cm
\includegraphics[width=0.36\textwidth]{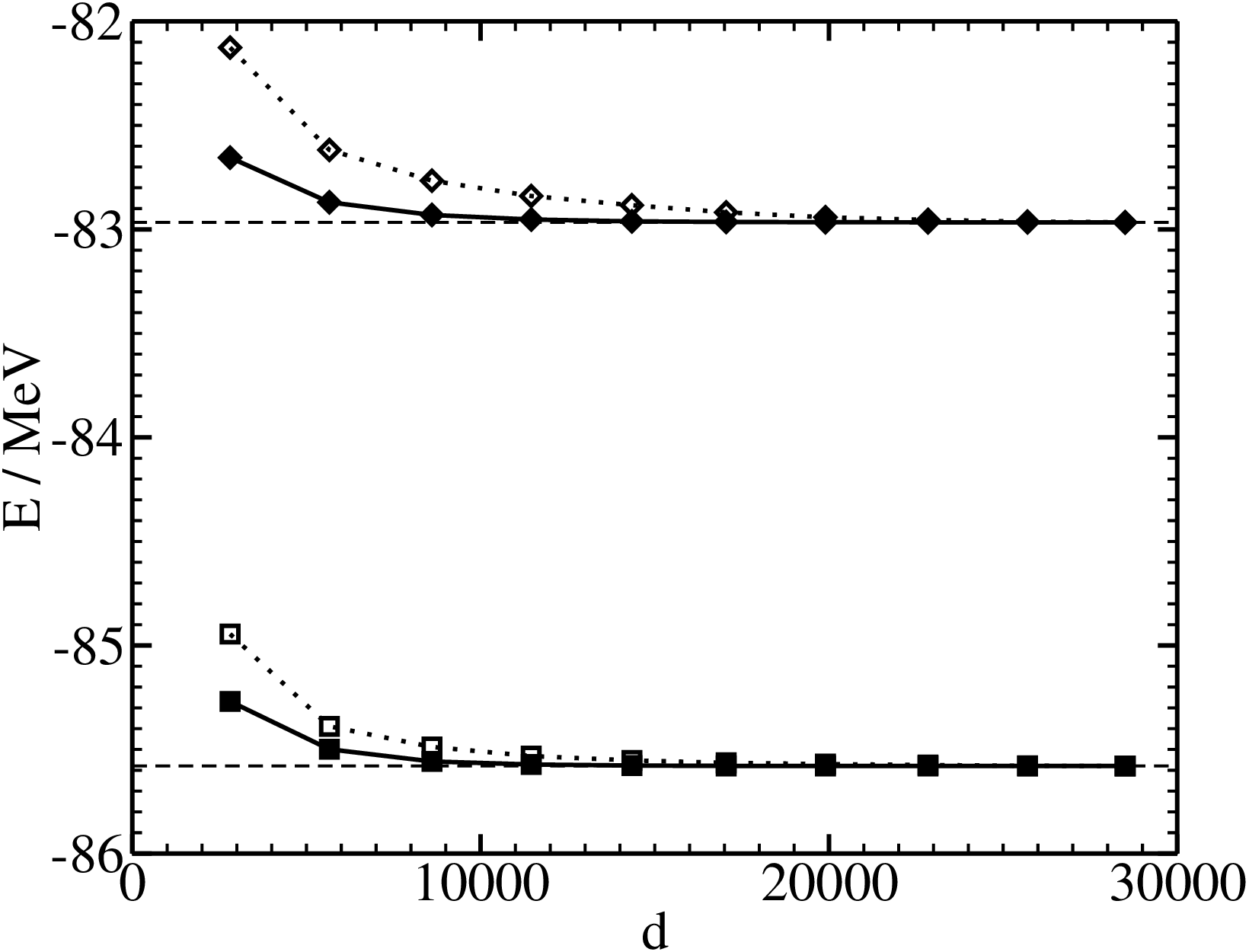}
\caption{\label{fig3}. Energies of the first and second excited 
states of $^{24}$Mg versus the dimension of the eigenvalue problem. 
Hollow data points: results from targeting the 
ground state. Filled data points: results from directly targeting the
excited states. Dashed lines: exact results.}
\end{figure}

\begin{figure}[b]
\vskip 0.3cm
\includegraphics[width=0.36\textwidth]{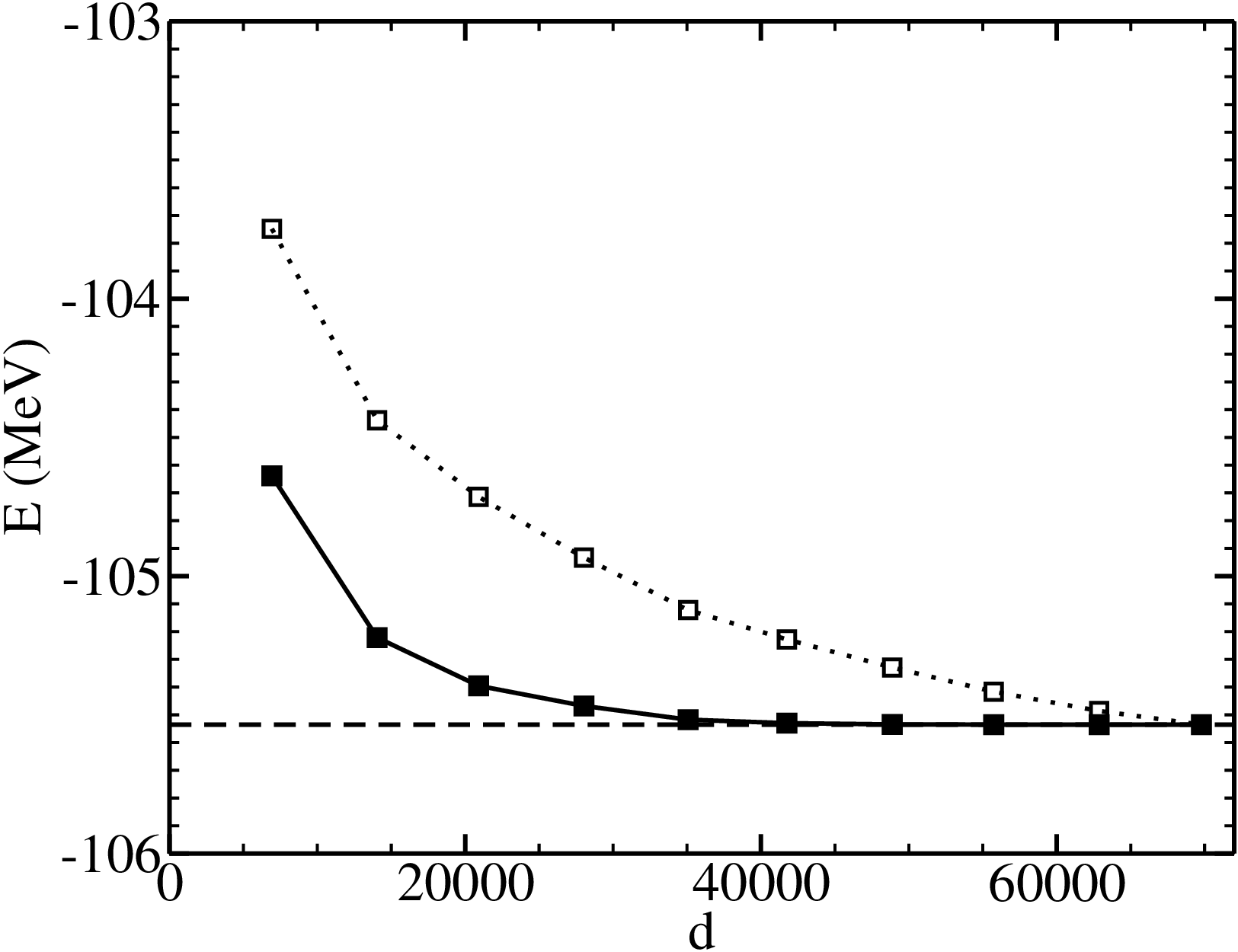}
\caption{\label{fig4}Energy of the first excited 
state of $^{26}$Al versus the dimension of the eigenvalue problem. 
Hollow data points: results from targeting the 
ground state. Filled data points: ground state calculation for
zero angular momentum. Dashed line: exact result.}
\end{figure}

So far we have only focused on the energies. In the remainder of this
section we discuss how the states and their quantum numbers are
reproduced by the factorization.  For ${}^{24}$Mg we analyze the
wave-function structure of the low-lying levels. Figure~\ref{fig5}
shows the squared overlaps with the exact results, obtained from
targeting the ground state. The solution of an eigenvalue problem of
only 10\% of the full dimension $d_{\rm max}$ already yields between
90-96\% of squared overlap. The directly targeted ground-state is
reproduced to more than 99\% once the dimension exceeds 20\% of the
full dimension $d_{\rm max}$.  The inset of Fig.~\ref{fig5} shows that
the defect $1-\langle\psi_{\rm exact}|\psi_{\rm factor}\rangle^2$
decreases exponentially fast as more factors are retained. While the
convergence is best for the directly targeted ground-state, the
excited states are also very accurately reproduced.

\begin{figure}[t]
\vskip 0.3cm
\includegraphics[width=0.36\textwidth]{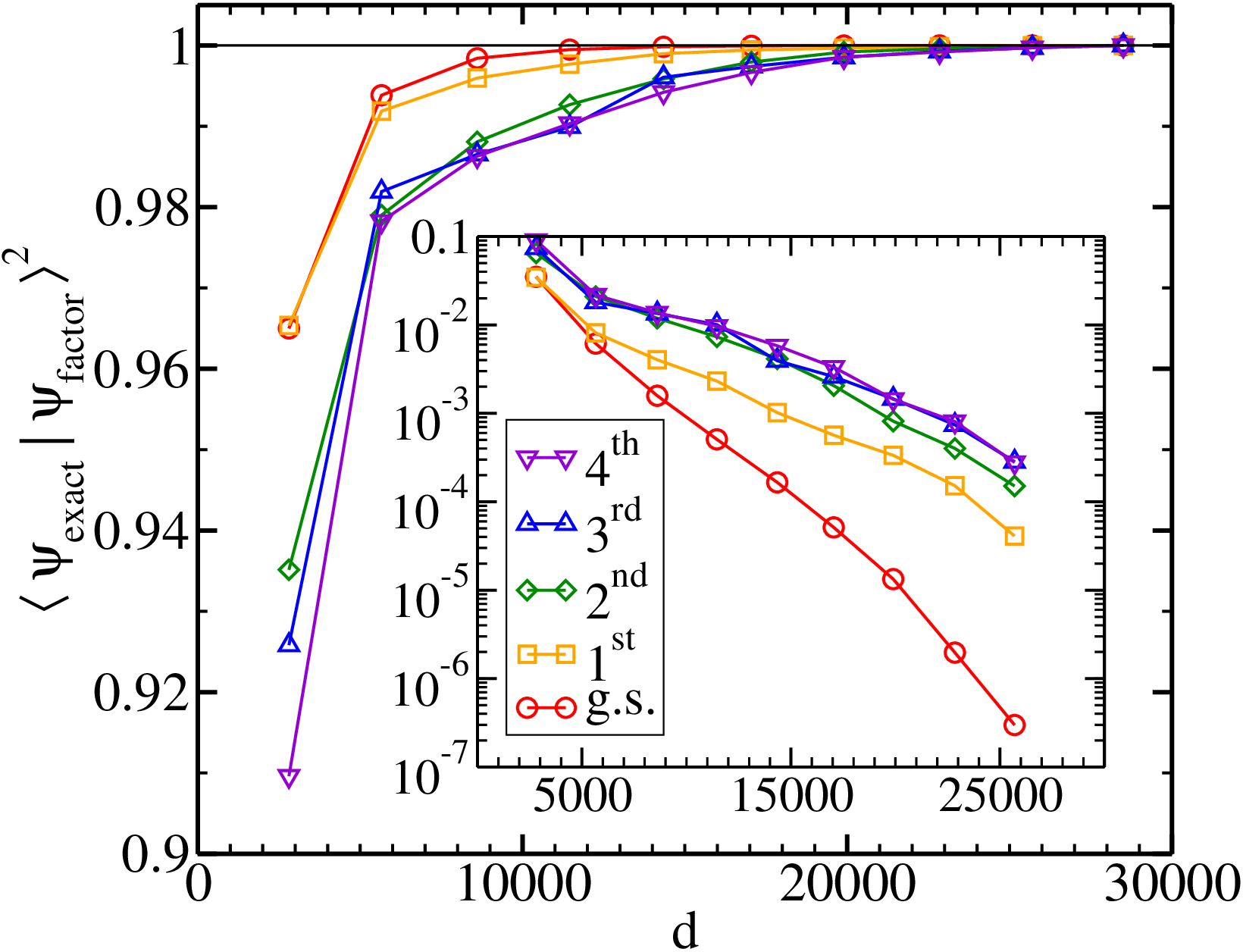}
\caption{\label{fig5}Squared overlaps $\langle \psi_{\rm
exact}|\psi_{\rm factor}\rangle^2$ for the low-lying states in
${}^{24}$Mg versus the dimension of the eigenvalue
problem (from targeting the ground state). 
The inset shows the deviation $1-\langle \psi_{\rm
exact}|\psi_{\rm factor}\rangle^2$ versus the dimension of the
eigenvalue problem.}
\end{figure}

Let us also verify for ${}^{24}$Mg that the quantum numbers of the
low-lying states are reproduced correctly. Due to rotational symmetry,
the expectation value for the angular momentum should fulfill 
\be
\langle \hat{J}^2\rangle=j(j+1), 
\ee 
where $j$ is a nonnegative integer. Figure~\ref{fig6} shows the
$j$-values of the low-lying states for ${}^{24}$Mg. The results were
obtained by targeting the ground state. The angular-momenta are very
accurately reproduced once about 20\% of the states are retained in
the factorization. This is remarkable since rotational symmetry is not
enforced in the $m$-scheme factorization, and no kind of constraint or
projection was used. Similar results are obtained for the total
isospin and for other $sd$-shell nuclei.  The accurate reproduction of
quantum numbers, wave-functions and energies implies that transition
matrix elements can accurately be computed.  This concludes or
detailed discussion of ${}^{24}$Mg.

\begin{figure}[t]
\vskip 0.3cm
\includegraphics[width=0.36\textwidth]{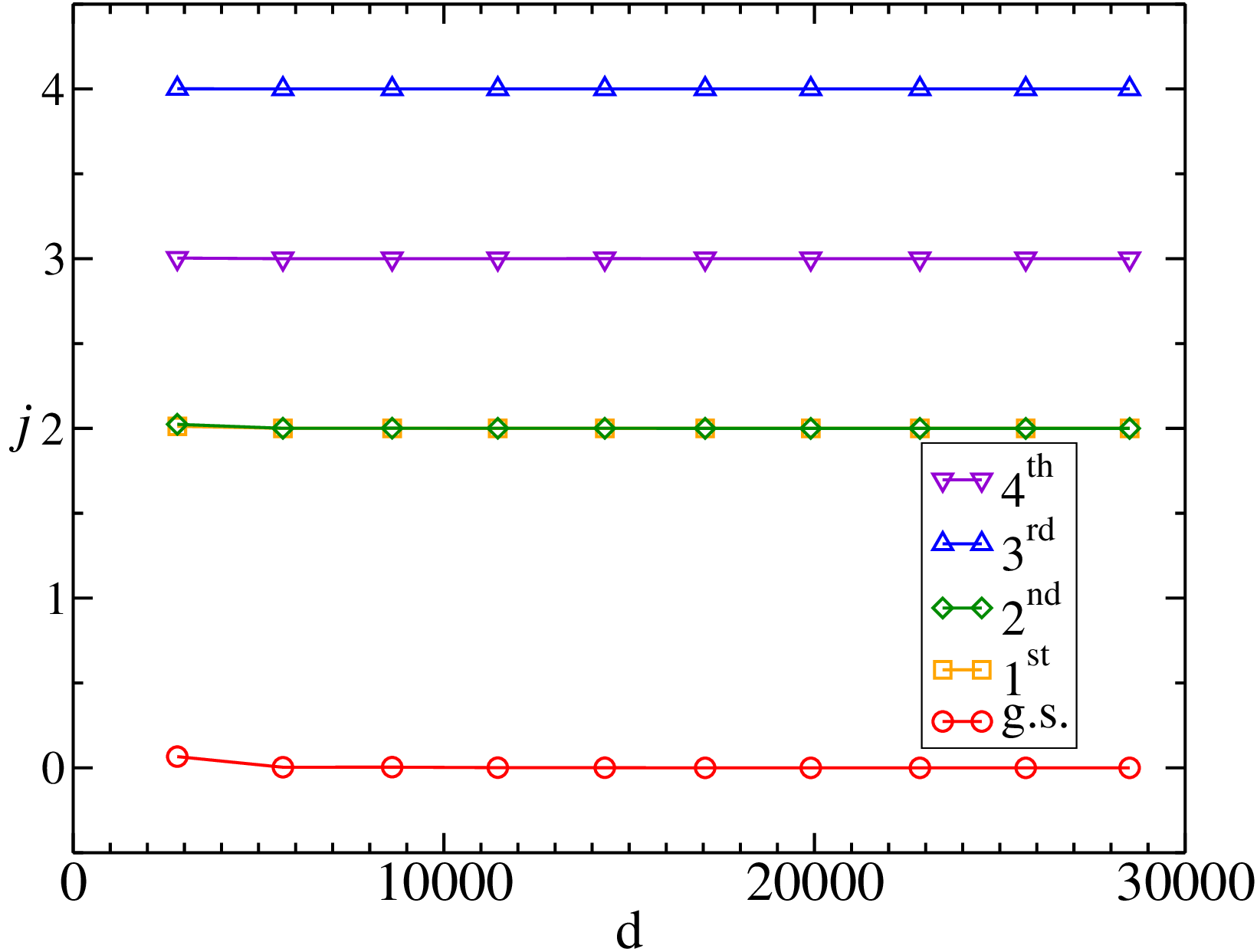}
\caption{\label{fig6}Angular momentum value $j$ for the low-lying states in
${}^{24}$Mg versus the dimension of the eigenvalue
problem (from targeting the ground state).}
\end{figure}

We finally mention that we also compared the $m$-scheme factorization
with the $jj$-coupled scheme. For $^{24}$Mg we find practically
identical convergence of the ground-state energy when plotted versus
the relative dimension $d/d_{\rm max}$ of the corresponding eigenvalue
problem. The dimensions $d$, $d_{\rm max}$ of the eigenvalue problem in the
$jj$-coupled scheme are, of course, smaller than for the
$m$-scheme. However, this does not translate directly into a
computational speed-up since the $jj$-scheme algorithm is more complex
and involves less sparse matrices. We believe that its main advantage
consists of the possibility to directly target the lowest state with a
given angular momentum.

\subsection{$pf$-shell nuclei}

Many theoretical results for $pf$-shell nuclei are available from
exact calculations for the KB3 interaction. In the lower
$pf$-shell, diagonalizations can be based on an $m$-scheme basis
\cite{Antoine}.  The $m$-scheme dimensions of upper $pf$-shell nuclei
exceed one billion, and exact diagonalizations have been performed in
a $J=0$ coupled basis \cite{Nathan}, reducing the dimensions to the
order of ten million. In this section we compare the results from
$m$-scheme factorization with the exact results. We are particularly
interested in the efficiency of the method and would like to answer
the following question. How does the relative dimension $d/d_{\rm
max}$, at which an accurate approximation to the ground state is
obtained, scale with increasing dimension $d_{\rm max}$ required for
an exact diagonalization?

\begin{figure}[b]
\vskip 0.3cm
\includegraphics[width=0.36\textwidth]{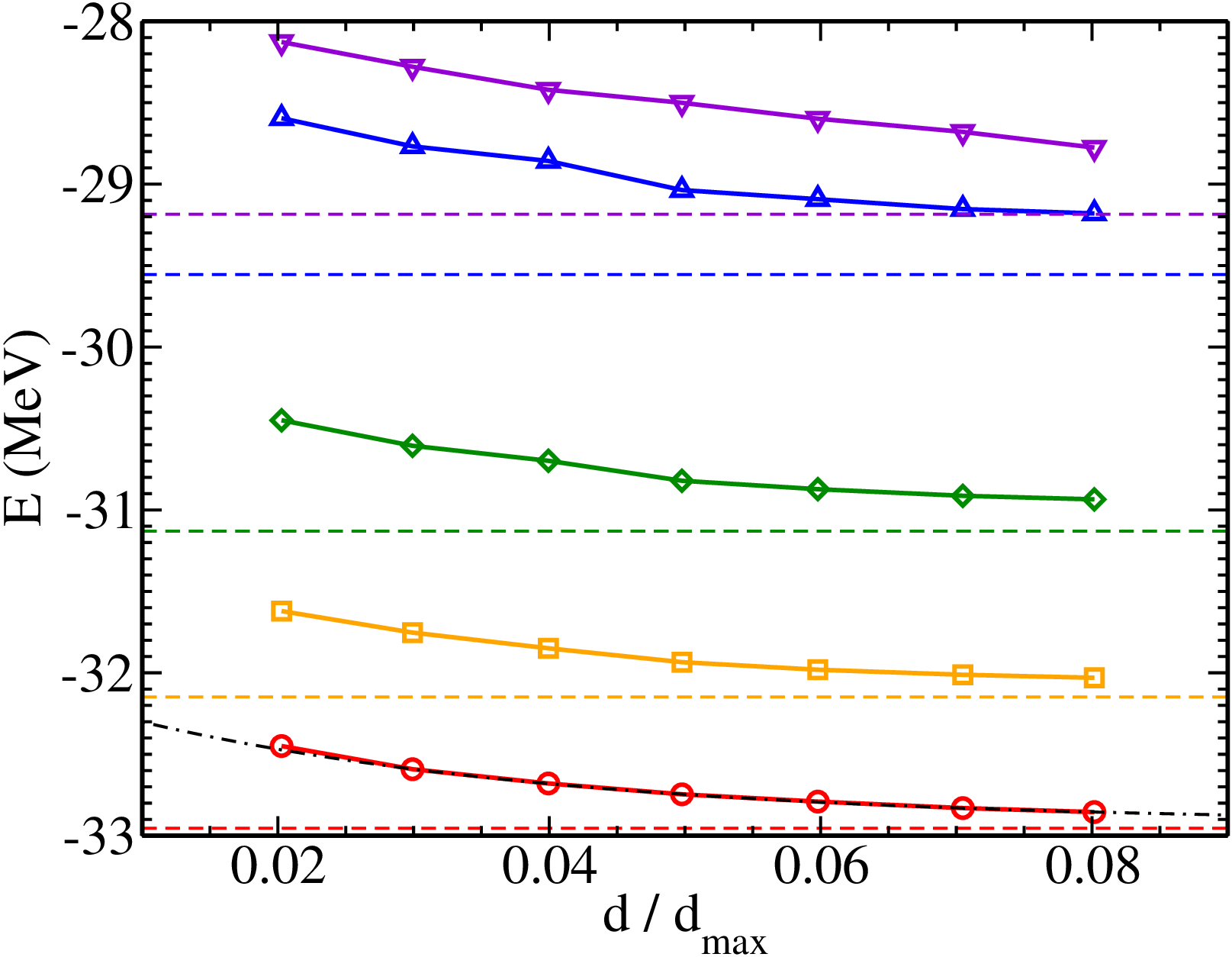}
\caption{\label{fig7}Data points: Low-lying energies of ${}^{48}$Cr
(KB3 interaction) versus the dimension $d$ of the eigenvalue problem
relative to the $m$-scheme dimension $d_{\rm max}$. Dashed lines:
Exact results. Dashed-dotted line: Exponential fit 
$E(d/d_{\rm max}) = -32.92 +0.851 \exp{(-31.38\,d/d_{\rm max})}$
to the ground state energy.}
\end{figure}

For $pf$-shell nuclei we use the KB3 interaction \cite{KB}.
Figure~\ref{fig7} shows the low-lying energies for ${}^{48}$Cr plotted
versus the relative dimension $d/d_{\rm max}$ of the eigenvalue
problem. The exact ground state energy is  $E_0=-32.95$ MeV
and results from the solution of an eigenvalue problem with dimension
$d_{\rm max}=1.96\times 10^6$ \cite{Antoine}.
The ground-state energy converges exponentially quickly as
the number of retained factors increases. The right-most data point
results from an eigenvalue problem with only 8\% of the $m$-scheme
dimension and involves $\Omega=\sum_m\Omega_m=391$ factors.  It
deviates less than 100 keV from the result of an exact
diagonalization. An exponential fit of the form $E(d/d_{\rm
max})=a+b\exp{(-cd/d_{\rm max})}$ to the right-most six data points
yields the estimate $E(1)=-32.92$ MeV, which is only 30 keV above the
exact result.  The excited states are obtained from
targeting the ground state.  The level spacings of the two lowest
excitations are very accurately reproduced even at the most severe truncation,
while the spacings to the higher levels are about 200-300 keV too
large. The angular momentum expectation values are plotted in
Fig.~\ref{fig8}. For $d/d_{\rm max} \agt 0.04$, energies and quantum
numbers are sufficiently well converged. Considering the modest size
of the eigenvalue problem we solved, these are very good results.

\begin{figure}[t]
\vskip 0.3cm
\includegraphics[width=0.36\textwidth]{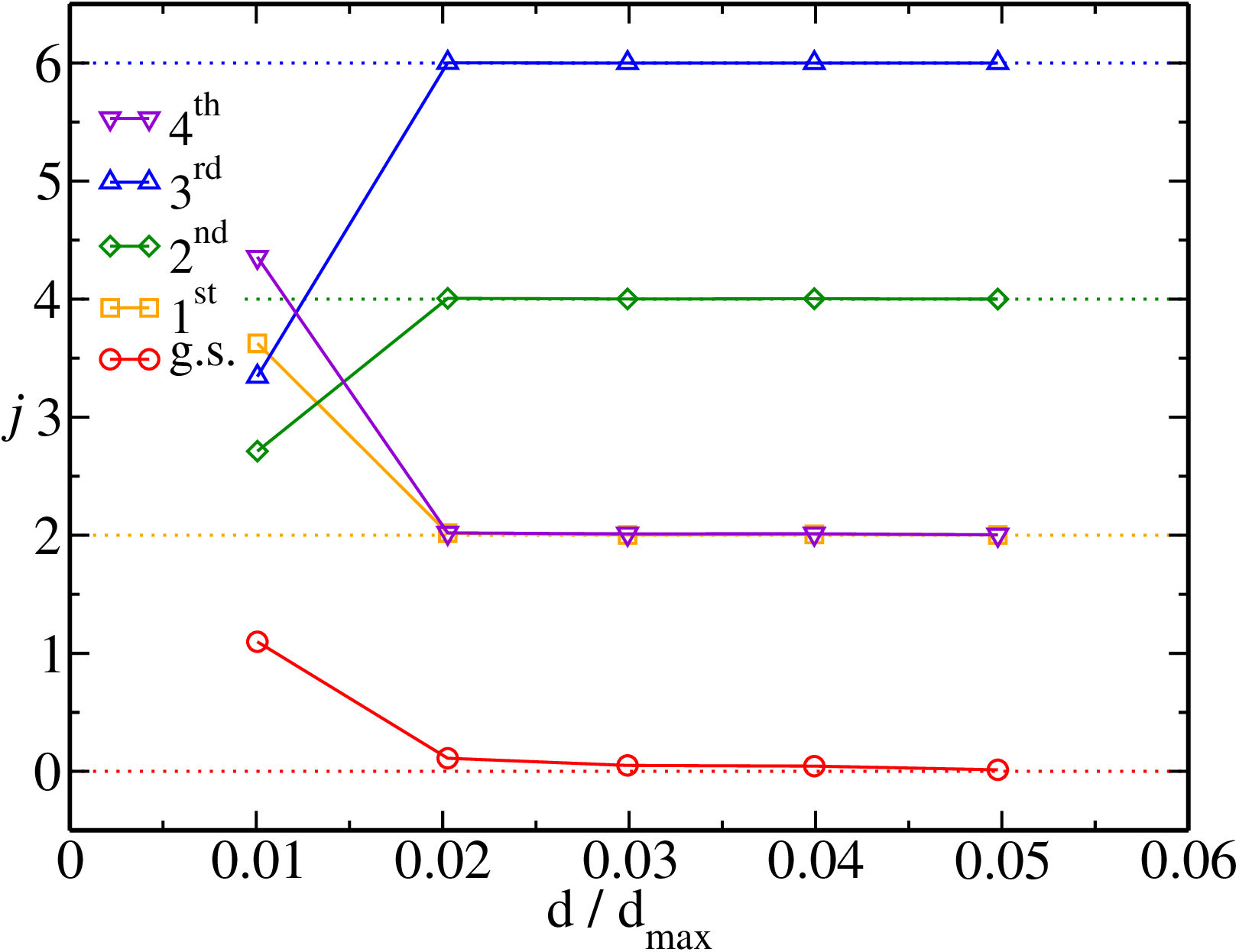}
\caption{\label{fig8}Angular momentum value $j$ for the low-lying excitations 
of the $pf$-shell nucleus ${}^{48}$Cr (KB3 interaction) plotted versus
the dimension $d$ of the eigenvalue problem relative to the $m$-scheme
dimension $d_{\rm max}$. The dotted lines are the exact results.}
\end{figure}

\begin{figure}[t]
\vskip 0.3cm
\includegraphics[width=0.36\textwidth]{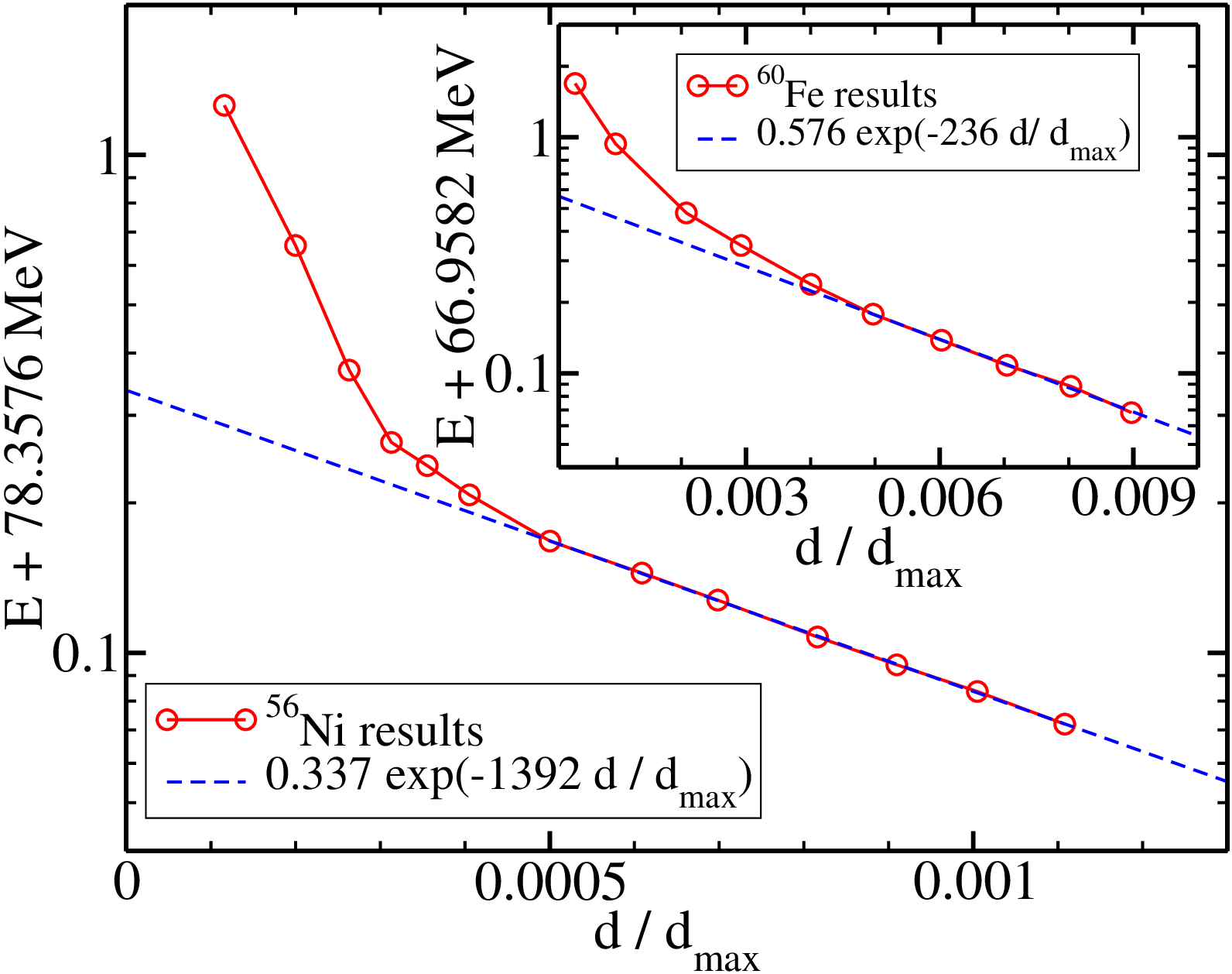}
\caption{\label{fig9}Ground-state energy $E$ versus the dimension $d$
of the eigenvalue problem relative to the $m$-scheme dimension
$d_{\rm max}\approx 1.09\times 10^9$ for $^{56}$Ni. The data points are from
the $m$-scheme factorization, and the dashed line is an exponential fit
to the data. Inset: Similar plot for
${}^{60}$Fe ($d_{\rm max}\approx 110\times 10^6$).}
\end{figure}

After this detailed discussion of $^{48}$Cr, we factor the ground
states of $^{60}$Fe and $^{56}$Ni. The exact energies are $E=-67.0$MeV
and $E=-78.46$MeV, respectively \cite{Nathan}, and the corresponding
$m$-scheme dimensions are $d_{\rm max}\approx 110\times 10^6$ and
$d_{\rm max}\approx 1.09\times 10^9$, respectively. Fig.~\ref{fig9}
shows that the ground states of these nuclei can very efficiently be
factored. Using an exponential fit to the numerical data points, the
ground state energies are reproduced within a deviation of 50 keV for
$^{60}$Fe and 100 keV for $^{56}$Ni. Most importantly, the relative
dimension of the eigenvalue problem we solve is $d/d_{\rm max} \approx
1\%$ for $^{60}$Fe (from $\Omega=\sum_m\Omega_m=352$ states) and about
$d/d_{\rm max} \approx 0.1\%$ for $^{56}$Ni (from
$\Omega=\sum_m\Omega_m=147$ states). This suggests that the
factorization gets increasingly efficient as the dimension of the
problem increases.

\subsection{No-core shell model}

As a final test, we consider a no-core shell-model problem and apply
the $m$-scheme factorization to $^{4}$He using a manageable model space and
a realistic interaction from a $G$-matrix calculation. The model space
consists of the 0$s$-0$p$-1$s$-0$d$-0$f$-1$p$ shells. The $G$-matrix
stems from a 15$\hbar\omega$ calculation and
is based on the Idaho-A potential \cite{IdahoA}. The Idaho-A potential
is derived from an effective Lagrangian that respects QCD inspired 
chiral symmetry. We calculate the $G$-matrix from 
\begin{equation}
G(\omega)=V + \frac{1}{\omega - QTQ}G(\omega)
\end{equation}
where $\omega$ is the starting energy, $T$ is the kinetic energy
operator, and $Q$ is the Pauli operator. Our Pauli operator allows for
all allowed configurations to be active in the chosen model space.  We
also employ folded-diagrams calculated at $\tilde{\omega}=-20.0$~MeV
to decrease the dependence of the resulting two-body interaction on
the starting energy. Details concerning the derivation of the
$G$-matrix may be found in Ref.~\cite{hj95}. We also include
center-of-mass corrections perturbatively.  Finally, we employ the
method of Ref.~\cite{dean99} to obtain an interaction that yields a
ground-state energy that is approximately free of center-of-mass
contamination.  Note that this small space is not sufficient to
completely describe the $^{4}$He nucleus, but still illustrates the
power of the factorization method for problems in which the core is
absent.

Figure~\ref{fig10} shows the energies for the three lowest lying
states of $^4$He versus the dimension $d$ of the eigenvalue problem we
solve. The results for the excited states were obtained while targeting
the ground state. The exact results are obtained from a diagonalization with
$m$-scheme dimension $d_{\rm max}=79298$. Note that the ground state
and the excited states converge very fast toward the exact results. 
A calculation with $d/d_{\rm max}\approx 0.2$ already yields excellent
approximations, and the angular momentum quantum numbers are converged. 

\begin{figure}[h]
\vskip 0.3cm
\includegraphics[width=0.4\textwidth]{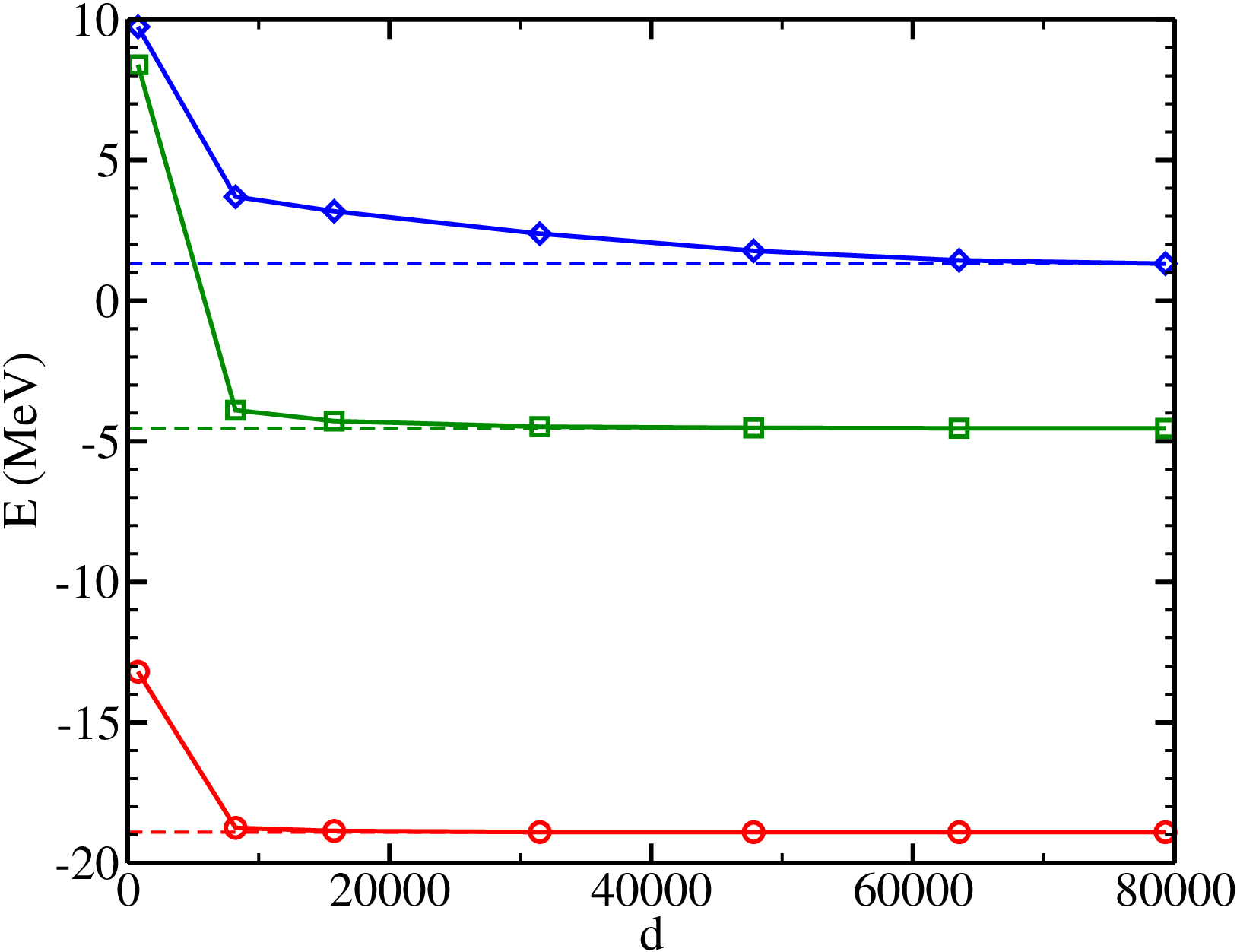}
\caption{\label{fig10}Energies of low-lying states of $^4$He
plotted versus the dimension $d$ of the eigenvalue problem.  The
dashed lines are the exact results.}
\end{figure}
 
Let us finally suggest an alternative treatment of the center of mass
problem.  The center of mass is separable in an oscillator basis where
all many-body states with up to $N_{\rm max}$ oscillator quanta are
included. The factorization could be applied in this scheme 
by combining proton states with $n$ oscillator quanta and neutron states
with $n'$ oscillator quanta such that $n+n'\le N_{\rm max}$.

\section{Convergence of the method} 
\label{conv}

\subsection{Convergence properties} 

The results of the preceding section showed that the factorization
converges exponentially quickly as more factors are retained.  So far
we considered nuclei with equal dimension of proton space and neutron
space, most of them being $N=Z$ nuclei. What can be expected for other
cases? To answer this important question, we computed the exact ground
states of several $pf$-shell nuclei and numerically performed singular
value decompositions of their amplitude matrices $\Psi_{\alpha \beta}$
as defined in Eq.~(\ref{state}). Fig.~\ref{fig11} shows logarithmic
plots of the resulting singular value spectra. The singular value
spectra exhibit a very sharp initial falloff followed by an
exponential decay. The initial falloff is stronger for larger
dimension of the proton space $d_P$, and this renders the
factorization method very effective.  There is no clear trend for
isotopic chains. The singular value spectra of the lighter $pf$-shell
nuclei decay most rapidly for the $N=Z$ nuclei, while the decay is
faster for mid-shell nuclei away from $N=Z$. Our results suggest that
the application of the factorization method is not limited to $N=Z$
nuclei. We also computed the number of factors, $\Omega(x)$, such that
$\sum_{j=1}^{\Omega(x)} s_j^2 > x$ for $x=0.99$ and $x=0.999$.  The
results of Table~\ref{tab2} show that the factorization of all these
nuclei should converge rapidly as more factors are included. Larger
dimensional model spaces usually require the retention of more
factors, but the ratio $\Omega(x)/\min{(d_P,d_N)}$ decreases with
increasing size of the problem. Note that $^{48}$Cr is relatively
difficult to factor into proton and neutron states. This suggests that
this nucleus exhibits particularly strong proton-neutron correlations.

\begin{figure}[t]
\vskip 0.3cm
\includegraphics[width=0.4\textwidth]{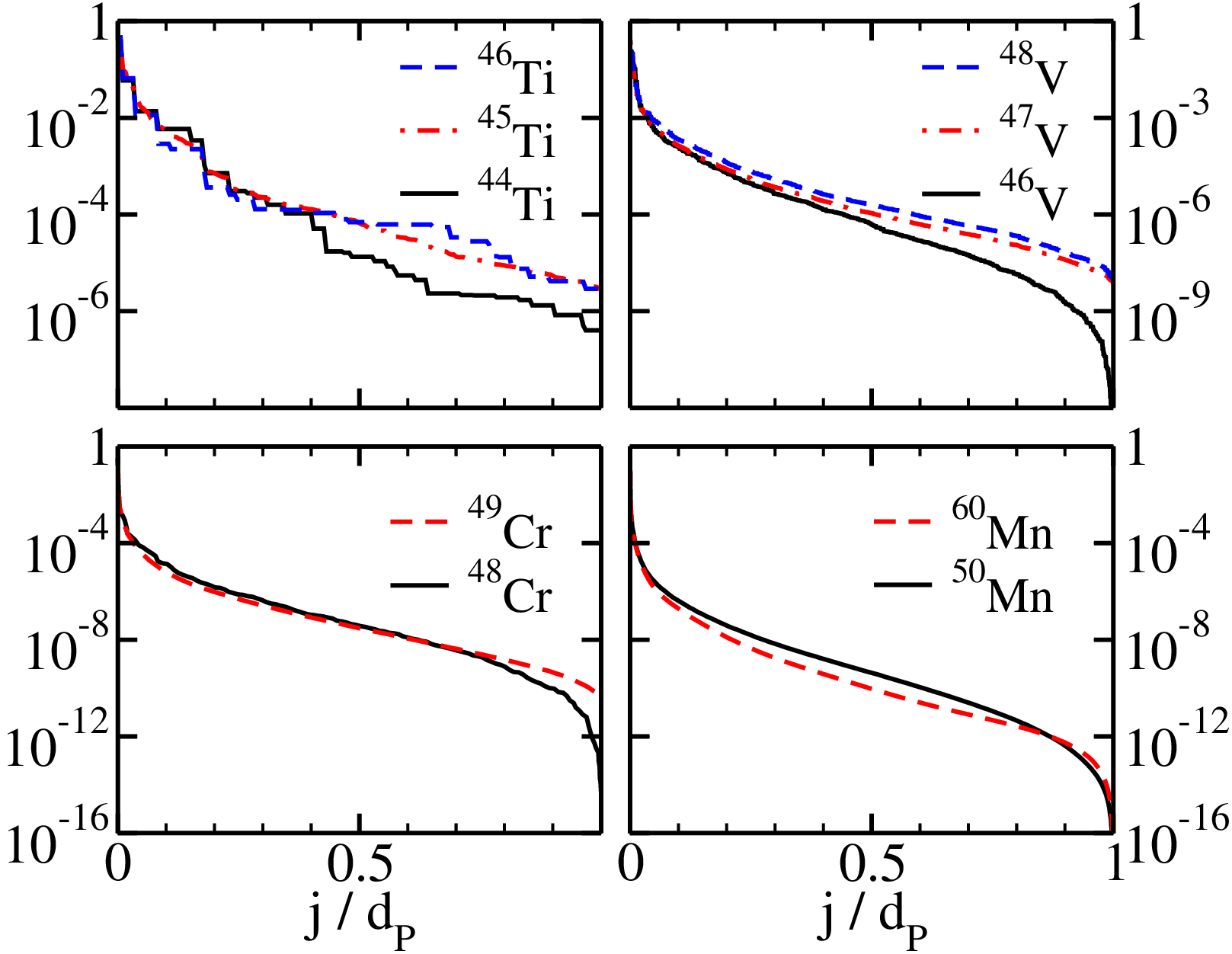}
\caption{\label{fig11}Singular value spectra (squared singular values
$s_j^2$) for ground states of Ti isotopes (upper left, $d_P=190$), V
isotopes (upper right, $d_P=1140$), Cr isotopes (lower left,
$d_P=4845$), and $^{50}$Mn (lower right, $d_P=15504$).  }
\end{figure}

\begin{table}[b]
\begin{ruledtabular}
\begin{tabular}{|c||r|r|r|r|}
Nucleus  & $d_P$  & $d_N$  & $\Omega(0.99)$ &$\Omega(0.999)$ \\\hline
$^{44}$Ti &   190  &   190 &  40 &  75 \\
$^{45}$Ti &   190  &  1140 &  45 & 109 \\
$^{46}$Ti &   190  &  4845 &  43 & 126 \\
$^{46}$V  &  1140  &  1140 &  98 & 270 \\
$^{47}$V  &  1140  &  4845 & 111 & 324 \\
$^{48}$V  &  1140  & 15504 & 151 & 392 \\
$^{48}$Cr &  4845  &  4845 & 258 & 775 \\
$^{49}$Cr &  4845  & 15504 & 168 & 619 \\
$^{50}$Mn & 15504  & 15504 & 197 & 821 \\
$^{60}$Mn & 15504  & 15504 & 163 & 570 \\
\end{tabular}
\end{ruledtabular}
\caption\protect{\label{tab2}
Proton space dimension $d_P$ and neutron space dimension $d_N$ for various
$pf$-shell nuclei. $\Omega(x)$ denotes the number of factors that have to
be retained for an overlap $\sum_{j=1}^\Omega s_j^2=x$ with the exact 
ground state.}
\end{table}

We recall that the $m$-scheme factorization (\ref{mscheme}) requires
as input the number of factors with a given angular momentum
projection, $\Omega_m$, which were taken according to
Eq.~(\ref{frac}). It is interesting and important to check this choice
of input parameters.  To this purpose we compare the singular value
spectrum from the factorization with the singular value spectrum from
an exact calculation.  The factorization was performed for $^{48}$Cr
using $\alpha=0.04$ and $\alpha=0.08$ in Eq.~(\ref{frac}). These
truncations included a total of $\Omega=\sum_m\Omega(\alpha)=197$ and
$\Omega=391$ factors, respectively. Figure~\ref{fig12} compares the
resulting singular value spectra with the singular value decomposition
of the exact ground state. The agreement between the exact results and
the approximations is rather good, and improves with increasing number
of retained factors. Note however, that the smaller singular values
deviate from each other. This suggests, that it should be possible 
to somewhat improve the choice of input parameters $\Omega_m$.

\begin{figure}[b]
\vskip 0.3cm
\includegraphics[width=0.4\textwidth]{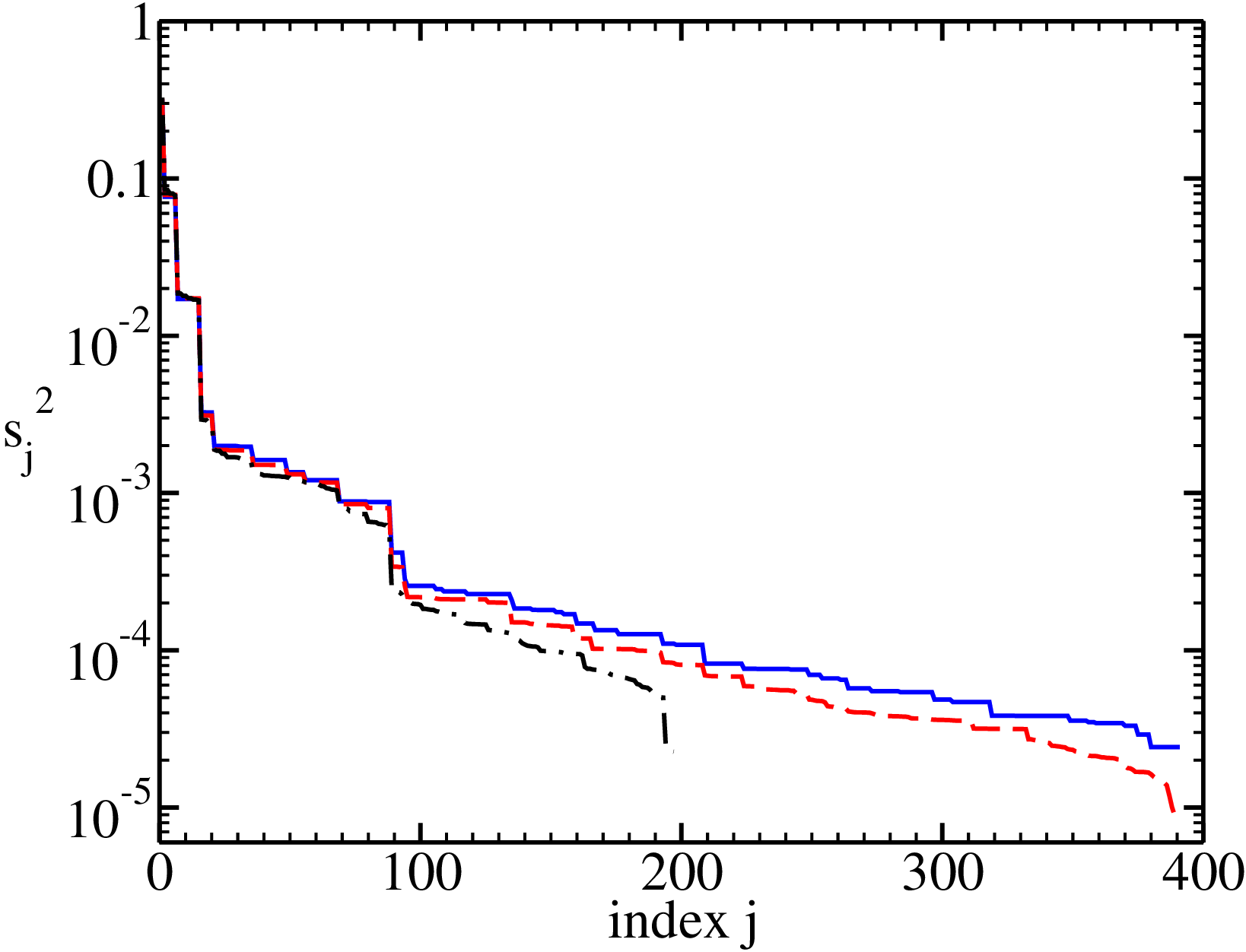}
\caption{\label{fig12}Singular value spectra (squared singular values
$s_j^2$) for  $^{48}$Cr. Singular value decomposition of exact
ground state (full line), $m$-scheme factorization using $\Omega=391$ factors
(dashed line), and $\Omega=197$ factors (dashed-dotted line).}
\end{figure}

We finally mention that we lack an understanding about the rapid
initial falloff in singular value spectra.  For the Hamiltonians
considered in this work the falloff is evident. Moreover, the results
obtained from DMRG calculations over the past decade demonstrate that
density-matrix spectra decay rapidly for the ground states of a large
number of relevant Hamiltonians. A few works address the theoretical
foundations of the DMRG. Peschel and coworkers investigated density
matrix spectra of several soluble problems \cite{Peschel}, while
Okunishi {\it et al.} discussed the asymptotic behavior of density
matrix eigenvalues for noncritical spin systems
\cite{Okunishi}. \"Ostlund and Rommer showed that if the DMRG
renormalization converges to a fixed point, the DMRG ground-state is
of a special matrix-product form \cite{Ostlund}. Given the lack of
generally valid analytical results, it is thus interesting to
numerically investigate singular value spectra for ``generic''
Hamiltonians. For this purpose we considered the model space of
$pf$-shell nuclei $^{44}$Ti and $^{46}$V and used a random two-body
interaction that preserves spin and isospin, i.e. the spin/isospin
coupled two-body matrix elements $V_{\alpha,\beta}$ are independent
Gaussian random variables with zero mean $\langle
V_{\alpha,\beta}\rangle=0$ and variance $\langle
V_{\alpha,\beta}V_{\alpha',\beta'}\rangle=
(\delta_{\alpha,\alpha'}\delta_{\beta,\beta'}+
\delta_{\alpha,\beta'}\delta_{\beta,\alpha'})$. For a realistic choice
of the single-particle energies we find singular value spectra that
are similar to the realistic spectra. However, setting all
single-particle energies to zero yields singular value spectra with
longer tails and a less rapid decay.

\subsection{Comparison with other truncation methods}

In recent years, several truncation methods have been developed for
and applied to shell model problems. In this subsection we compare
some of these approaches with the method presented in this work. This
comparison focuses on convergence and accuracy of low-lying
energy spectra at a given level of truncation. 

We start with the DMRG which also bases its truncation on the singular
values \cite{White92}. Dukelsky and coworkers applied the DMRG to
nuclear structure problems involving pairing \cite{Duk01} and
pairing-plus-quadrupole interactions \cite{Duk02}.  These applications
were very successful as accurate results could be obtained for huge
Hilbert spaces. The factorization method proposed in this work can
only treat much smaller Hilbert spaces and cannot compete with the
DMRG for these systems. However, the recent DMRG
calculation~\cite{Dimitrova02} for the realistic nuclear structure
problem ($^{24}$Mg with USD interaction) converges very slowly.

Recently, Andreozzi {\it et al.} \cite{AP,ALP} used a small number of
correlated proton states and correlated neutron states as a truncated
basis for shell model problems.  The correlated proton (neutron)
states are the low-energy eigenstates of the proton-proton
(neutron-neutron) Hamiltonian. The full shell model Hamiltonian
including the proton-neutron interaction is then solved in this space.
Andreozzi and Porrino report exponentially converging results and a
considerable reduction in the number of basis states \cite{AP}. This
procedure differs from our approach mainly by the absence of a
variational principle.

A third related method is the exponential convergence method (ECM)
developed by Horoi and coworkers
\cite{Horoi94,Horoi99,Horoi02,Horoi03}. In this method, shell-model
configurations are ordered according to their average centroid, which
are obtained from statistical spectroscopy. This ordering gives a
natural truncation scheme, and analytical arguments suggest an
exponential convergence of energies with increasing number of retained
configurations. Once the exponential region is identified, the full
space energies can be extrapolated by an exponential fit. A direct
comparison is not easy since the FPD6 interaction is used for
$pf$-shell nuclei, and since ECM results are plotted versus
$JT$-coupled dimension of the truncated space. We believe, however,
that our method converges more rapidly than the ECM. For $^{48}$Cr,
for instance, our rate of exponential convergence is $c\approx -31.38$
(See Fig.~\ref{fig7}), which is about a factor eight larger than what
is reported for the ECM in Fig.~1 of Ref.~\cite{Horoi02}. For
$^{56}$Ni, our exponential rate is about a factor 200 larger than the
ECM rate \cite{HoroiPC}, and our identification of the exponential
region requires a $m$-scheme dimension $d\approx 10^6$ (See
Fig.~\ref{fig10}) while the ECM requires an $m$-scheme dimension of
4-5 million \cite{HoroiPC}.

For upper $pf$-shell nuclei, truncations can be based on the maximal
number $t$ of nucleons outside the $f_{7/2}$ subshell
\cite{Caurier}. Within this truncation, the convergence of the energy
is rather slow, and it is difficult to extrapolate from results in
truncated spaces to the full Hilbert space. Mizusaki and Imada devised
extrapolation methods that link the error due to the truncation to the
variance of the energy in the truncated space. This approach lead to a
first order \cite{Mizu02} and a second-order extrapolation method
\cite{Mizu03} for predictions of low-lying states in various
$pf$-shell nuclei. For $^{56}$Ni the approximation of a closed
$f_{7/2}$ subshell is well justified, and the
extrapolation methods yields results that are superior to the
factorization \cite{MizusakiPC}.  For $^{48}$Cr, however, the
factorization seems to be of advantage: The exact ground-state
energy being $E=-32.95$ MeV and the $m$-scheme dimension $d=1.96\times
10^6$. The first-order extrapolation method \cite{Mizu02} yields
$E=-33.008$ for $t=5$ and $E=-32.975$ for $t=6$, and the corresponding
$m$-scheme dimensions are $d(t=5)\approx 1.3\times 10^6$ and
$d(t=6)\approx 1.76\times 10^6$. The second-order extrapolation
(``scheme I'') \cite{Mizu03} yields $E=-32.91$ for $t=5$.  Slightly
better results are obtained from a different truncation scheme
(``scheme II''). The ground-state factorization yields a comparably
good energy estimate $E=-32.92$ MeV (See Fig.~\ref{fig7}) from solving
a much smaller eigenvalue problem of dimension $d=1.6\times 10^5$.

The mixed-mode shell model approach developed by Gueorguiev {\it et
al}~\cite{Vesselin01,Vesselin02} combines single-particle
configuration and $SU(3)$ collective configurations to describe the
interplay and competition between single-particle and collective
degrees of freedom. For the $sd$-shell nucleus $^{24}$Mg, the
mixed-mode shell model yields good approximations to the binding
energy (within 2\% deviation of the exact result), and low-energy
configurations which exceed 90\% overlap with the exact results. These
results stem from a truncated space of only 10\% the full Hilbert
space \cite{Vesselin02}. At the 10\% level of the truncation, the
factorization method yields an energy deviation of less than 1\% (See
Fig.~\ref{fig2}), and squared overlaps exceed 96\% for the two lowest
lying states and 90\% for the following three states (See
Fig.~\ref{fig5}).

The method proposed in this work thus compares well to most of the
alternatives regarding convergence and accuracy at a given level of
truncation. Note, however, that its implementation seems somewhat more
complex than a shell-model approach with a configuration truncation
and somewhat less complex than the DMRG algorithm.

\section{Conclusion}
\label{concl}

We approximated the ground states of realistic nuclear structure
Hamiltonians by sums over products of correlated proton states and
correlated neutron states. The optimal states are determined by a
variational principle and are the solution of rather low-dimensional
eigenvalue problems. Computations for $sd$-shell nuclei, $pf$-shell
nuclei, and no-core shell models show that the method converges
exponentially quickly as more factors are included, and that accurate
approximations to shell-model ground states and low-lying excitations
may be obtained. For the largest problems we considered, the dimension
of the eigenvalue problem was reduced by three orders of
magnitude. Momentarily, the application of this method is limited by
the size of the proton space and the neutron space. An interesting
future development would also consider the factorization of these
spaces in order to treat larger dimensional problems. While the reason
of the exponential convergence is not yet understood, computations of
shell model problems with realistic and random two-body interactions
suggest that this behavior can be expected for a variety of
interactions.

\section*{Acknowledgments}

The authors acknowledge useful discussions with M.~Horoi, N.~Michel,
T.~Mizusaki and G.~Stoitcheva, and thank M.~Hjorth-Jensen for
providing us with a $G$-matrix.  This research used resources of the
Center for Computational Sciences (Oak Ridge National Laboratory) and
the National Energy Research Scientific Computing Center
(Berkeley). This work was supported in part by the U.S. Department of
Energy under Contract Nos.\ DE-FG02-96ER40963 (University of
Tennessee) and DE-AC05-00OR22725 with UT-Battelle, LLC (Oak Ridge
National Laboratory).

\end{document}